\documentclass[pre,aps,preprint, superscriptaddress]{revtex4-1}
\pdfoutput=1
\usepackage{amsmath,natbib,amsfonts}
\usepackage{amssymb,color, float}
\usepackage{graphicx}
\usepackage[centerlast]{caption}
\usepackage{subfig}
\usepackage{tikz}
\usepackage[colorlinks=true, citecolor=blue, urlcolor = blue, linkcolor= red,bookmarks=true]{hyperref}
\newcommand{\beqa}{\begin{eqnarray}}
\newcommand{\eeqa}{\end{eqnarray}}

\begin{document}
\title{Exactly solvable model of a passive Brownian heat engine and its comparison with active engines.}
\author{Rita Majumdar} 
\affiliation{Department of Physics, Indian Institute of Technology Delhi, New Delhi-110016, India.}
\author{Arnab Saha}
\affiliation{Department of Physics, University Of Calcutta, 92 Acharya Prafulla Chandra Road, Kolkata-700009, India}
\email{sahaarn@gmail.com}
\author{Rahul Marathe}
\email{maratherahul@physics.iitd.ac.in}
\affiliation{Department of Physics, Indian Institute of Technology Delhi, New Delhi-110016, India.}
\date{\today}
\begin{abstract}
  We perform an extensive analysis of passive as well as active micro-heat engines with different single-particle stochastic models. Using stochastic thermodynamics we calculate thermodynamic work, heat, entropy production and efficiency of passive and active Brownian heat engines analytically as well as numerically and compare them. We run the heat engines with a protocol for which the average thermodynamic quantities are calculated exactly for an arbitrary cycle time. We also discuss about the group of protocols for which exact non-quasistatic calculations can be done, completely in the passive engine case and partially in the active engines. We obtain detailed thermodynamics of non-quasistatic (i.e. powerful) single-particle micro heat engines. The quasistatic (i.e. zero power) limit of the results is obtained by taking long (infinite) cycle time. We also study the distributions of position of the confined particle in both passive and active engines. We compare their characteristics in terms of the parameter that measures the competition between the active persistence in the particle position (due to active noises) and the harmonic confinement. We also calculate excess kurtosis that measures the non-Gaussianity of these distributions. Our analysis shows that  efficiency of such thermal machine can be enhanced or reduced depending on the activity present in the model.
  \end{abstract}
\maketitle  

\section{Introduction}

Microscopic thermal devices operating under large thermal fluctuations is a frontier field of research \cite{Martinez17, Martinez16,Seifert12, Gracia16,Marathe07,  Basu17, Rana14}. They convert ambient thermal energy into mechanical work. They can extract heat from one reservoir of higher temperature, convert it partially to mechanical work and rest they deposit into another reservoir of lower temperature, similar to popular macroscopic engines like Carnot or Stirling. In the quasistatic or zero power limit, the efficiency of the microscopic engines is bounded by Carnot efficiency \cite{Seifret05, Seifert08, Sekimoto98, Seifert18}, and at maximum power the efficiency is bounded by Curzon-Ahlborn efficiency \cite{Curzon, Tschmidl08, Vholubec18, Johal}. Though, in contrast to the macroscopic devices, in the microscopic systems, due to their minute size, the magnitude of energy fluctuation is comparable to the average energy flows into or out of the device. Therefore, the performance of such machines strongly depend on the fluctuations present in their environment. For such small systems, even if driven arbitrarily far from equilibrium \cite{Fodor16,Dabelow}, thermodynamic quantities like work, heat, efficiency, entropy etc. can be defined and estimated along individual trajectories of the system within recently developed theoretical framework of stochastic thermodynamics \cite{Tschmidl08, Sekimoto98, Seifert12, Feynman64, Seifert08, Seifret05, Sekimoto10}. Motivated by experiments done with tunable laser traps where a single colloidal bead can be confined, theoretical models of single-particle microscopic thermal devices, such as heat engines, refrigerators and heat pumps are being developed \cite{Bechinger12, Volpe12, Robnagel, Rana16, JungWan21}.

In the last decade or so, this research area is being extended to systems where mechanical work may be extracted not only from thermal fluctuations but also from athermal fluctuations, in particular, from {\it{active}} fluctuations \cite{Krishnamurty16, Arnab18, Arnab20, Mandal17, Viktor17, Arnab19, Rana, maggi14, Zakine17, FodorHaya18, Goswami19, Slahiri20, Gomez21, Cates21, Cates21EPL, Cates22, Nroy21}.
The typical model systems that have been realised both experimentally and theoretically, are again a single colloidal bead trapped within an externally tunable optical potential and immersed in the active reservoir (e.g. - bacterial suspensions or active colloidal suspensions \cite{Leonard10, Krishnamurty16, Fodor20}), unlike the passive microscopic thermal devices where passive thermal baths are used. It has been shown experimentally that using bacterial suspensions as a non-equilibrium reservoir \cite{Krishnamurty16} and theoretically with simple model systems, that the performance of such {\it microscopic thermodynamic devices} can exceed their passive counterparts \cite{Martinez16, Fodor16, Krishnamurty16, Arnab18, Arnab20}. It has been shown recently that activity in the reservoir can cause more dissipation \cite{Steffenoni} or induce colour into the fluctuations of the dynamics of the colloidal bead, breaking fluctuation dissipation relation (FDR) and driving the system away from equilibrium \cite{Viktor20, Ramaswamy10, Szamel19, Fodora18, Fodor16}.      

In most of the above calculations, the average efficiency is estimated in the quasistatic (i.e. zero power) limit. Here, we report a simple  model of micro heat engine using a passive colloidal particle as the working substance, which is trapped in a harmonic potential. The stiffness of the trap is varied periodically over time using a specific class of protocol such that all the average thermodynamic quantities (e.g. work, heat, entropy production, efficiency) can be calculated exactly for all time-periods (or cycle times) of the protocols. The protocol follows the Stirling cycle. Apart from efficiency at quasistatic limit (i.e. at zero power limit), this will also allow us to calculate the efficiency at maximum power. Using such protocols we calculate average thermodynamic quantities in case of passive as well as active micro heat engines. We consider two different types of activities that the working substance (passive colloidal particle) encounters due to the reservoir: (i) active particles in the reservoir following run-and-tumble (RT) dynamics \cite{Slahiri20, Zakine17, Chaki19} and  (ii) active particles in the reservoir following Active Ornstein–Uhlenbeck process (AOUP) \cite{Arnab18, Arnab20, Fodor18}. 

We will compare the performance of these three types (passive, AOUP and RT) of microscopic thermal devices by calculating them analytically as well as numerically.

In what follows we first describe the models of micro-heat engines. Next we deduce the generic expressions for the average thermodynamic quantities with passive as well as active reservoirs. Then we move to a special case where a particular class of protocol will be introduced to compute the average thermodynamic quantities exactly for both passive and active cases. Then the thermodynamics of passive and active systems is compared. Finally we summarize and conclude.

\section{Model}
\begin{figure}[!t]
\begin{center}
\includegraphics[width=10cm,height=8cm,angle=0]{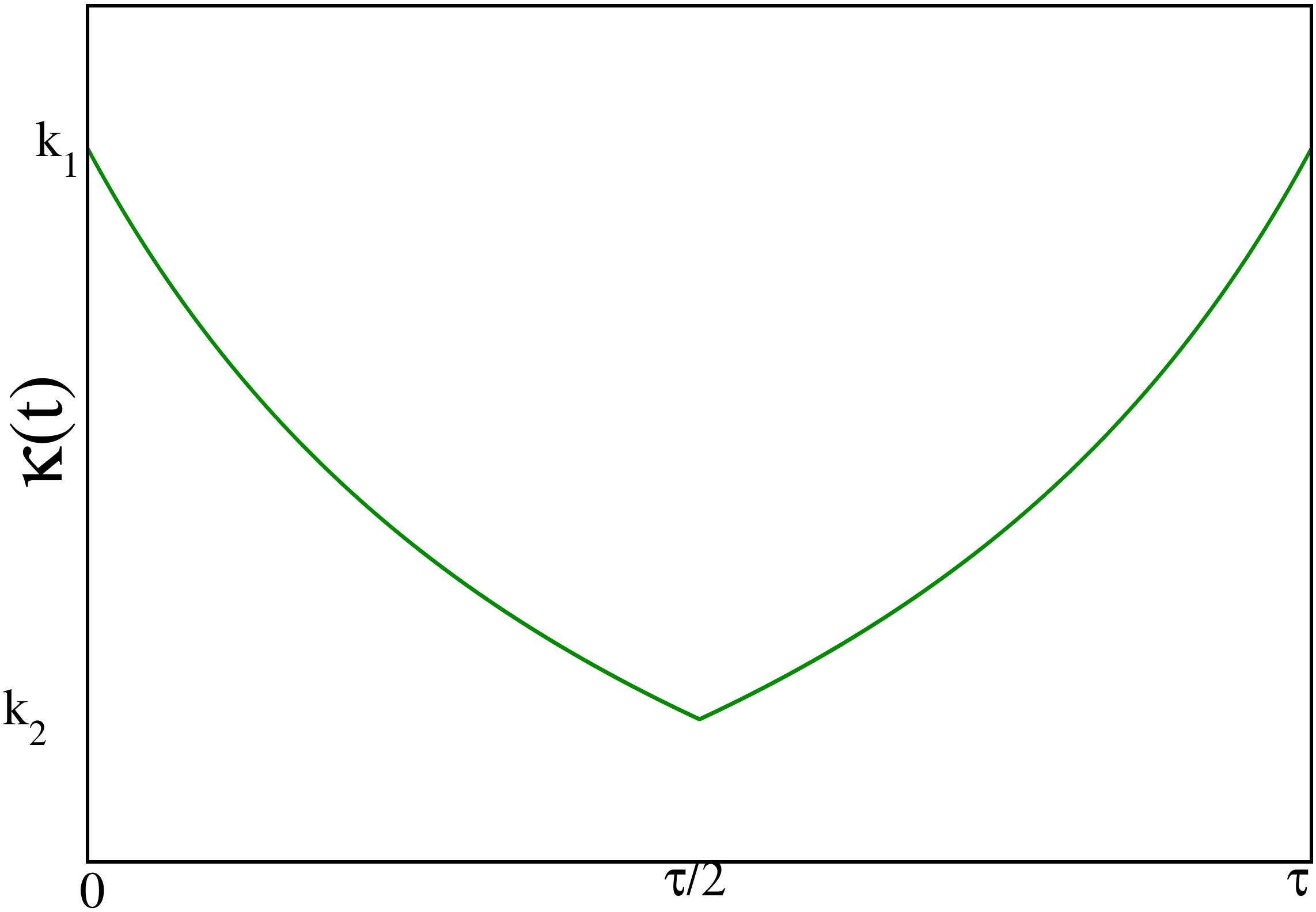}
\caption{A particular type of the time dependent protocol of the stiffness $\kappa(t)$. $k_1$ and $k_2$ are the maximum and minimum values of the trap strength, and $\tau$ is the cycle time.} 
\label{Fig.1}
\end{center}
\end{figure}

We consider a one dimensional system that consists of a colloidal particle at position $x(t)$ at time $t$, confined by a harmonic trap with  time-periodic and continuous stiffness $\kappa(t,\tau)$ where $\tau$ is the time period of a cycle. We consider $\kappa(t,\tau)$ to be a decreasing function in the first half of the cycle ($0 \leq t \leq \tau/2$) mimicking isothermal expansion and an increasing function in the second half of the cycle ($\tau/2 \leq t \leq \tau$) mimicking isothermal contraction (see Fig. \ref{Fig.1}). One complete cycle of duration $\tau$ mimics a Stirling like protocol. As the stiffness is continuous and periodic in time, $k_1=\kappa(t=0)=\kappa(t=\tau)$ and $k_{2}=\kappa(t=\frac{\tau}{2})$. $(k_1, k_2)$ are the maximum and minimum value $\kappa(t,\tau)$ respectively. The particle is suspended in a bath which has passive as well as active components. The passive degrees of freedom are equilibrated at temperature $T_{1}$ in the first half and at temperature $T_{2} < T_1$ in the second half of the cycle. We assume the transition from $T_{1}$ to $T_{2}$ at the time $t=\tau/2$ to be instantaneous. In such a case the equation of motion of the colloidal particle is given by the over-damped Langevin equation \cite{Sekimoto98, Seifert08}
\begin{equation}
\gamma_{i}\dot{x}_i(t)=-\kappa(t)x_i(t)+\xi_{i}(t)+\eta_{i}(t),
\label{eom}
\end{equation}
where $i = 1, 2$ corresponds to the first and the second half of the cycle respectively and $(^.)$ corresponds to the single time derivative. Here $\xi_i(t)$'s are the thermal noises and $\eta_{i}(t)$'s the active noises in the two halves of the cycle. The thermal noises are the standard Gaussian white noises such that
\beqa
\langle \xi_{i}(t)\rangle =0, ~\langle\xi_i(t)\xi_i(t')\rangle~=~2\gamma_i k_B T_{i}~\delta(t-t'),~ i=1,2,
\label{noise}
\eeqa
we take $\langle \eta_i(t)\rangle =0$ as well. We note that due the presence of the active fluctuations (that break the FDR) this system is inherently non-equilibrium. We will discuss the details of the particular form of active noises in the section \ref{ActivEngs}. In a purely passive case ($\eta_i(t)=0$), we obtain the well studied passive microscopic thermal devices including engines, refrigerators and pumps \cite{Tschmidl08, Rana, Rana16}. The formal solution of Eq. (\ref{eom}) is given as  
\begin{equation}
x_i(t)=\exp\left(-\frac{1}{\gamma_i}\int^t{dt^{\prime}\kappa(t^{\prime})}\right) \left[x_{0i} +\frac{1}{\gamma_i}
    \int^t {dt^{\prime}~(\xi_i(t')+\eta_i(t'))}\exp\left(\frac{1}{\gamma_i}\int^{t'}{dt''\kappa(t'')}\right) \right].
\label{formal}
\end{equation}
If we consider a particular protocol namely
\beqa
\frac{1}{\gamma_i}\int^t{dt^{\prime}}\kappa(t^{\prime}) &=& \ln f_i\left(t\right),
\label{protocol1}
\eeqa  
where $f_i(t)$ is a dimensionless function of time. For any regular function $f$,  Eq. (\ref{protocol1}) defines a class of all the protocols for which  $\kappa(t,\tau)\sim \frac{\dot f}{f}$.  Using this the average thermodynamic quantities can be calculated exactly at least for the passive case. For such a protocol Eq. (\ref{formal}) can be written as 
\begin{equation}
x_i(t)=\frac{x_{i}(0)}{f_i(t)} + \frac{1}{\gamma_if_i(t)}\int^t dt'~\left(\xi_i(t')+\eta_i(t')\right)~f_i(t').
\label{formal1}
\end{equation} 
The functional form of $f_1(t)$ and $f_2(t)$ can be found out after integrating $\kappa(t)$ in two halves of the cycle. In the first half the integration will run from $0$ to $t\leq \tau/2 $ and in the second from $\tau/2$ to $t\leq\tau$. In general the values of the functions $f_1(t=\tau/2) $ and $f_2(t=\tau/2) $ may not match. 

\section {Stochastic Thermodynamics}
In this section we define different thermodynamic quantities required for our analysis. From Eq. (\ref{eom}) one can write 
\begin{equation}
\dot x_i(t)(\gamma_i\dot x_i(t)-(\xi_i(t)+\eta_i(t)))=-\frac{d}{dt}\left [\frac{1}{2}\kappa(t)x_i^2(t)\right]+\frac{1}{2}\dot \kappa(t) x_i^2(t).
\label{firstlaw}
\end{equation}
Now, if we identify the trajectory dependent heat as $q_i(t)\equiv\int dt\left(\dot x_i(t)(\gamma_i\dot x_i(t)-(\xi_i(t)+\eta_i(t)))\right)$, work as $w_i(t)\equiv\int dt \frac{1}{2}\dot\kappa(t)x_i^2(t)$ and internal energy $u_i\equiv\frac{1}{2}\kappa(t)x_i^2$ associated with the particle \cite{Sekimoto98, Seifert08}, Eq. (\ref{firstlaw}) can be written as
\begin{equation}
\frac{dq_i(t)}{dt}=-\frac{du_i(t)}{dt}+\frac{dw_i(t)}{dt}.
\label{1stlaw}
\end{equation}
We identify Eq. (\ref{1stlaw}) with the first law of stochastic thermodynamics of a Brownian particle suspended in an active bath. Below we motivate the identification of various thermodynamic quantities. We note that the working substance, i. e. the Brownian particle is itself passive, trapped by a harmonic potential and follows an over damped dynamics. Therefore we relate the first term on the RHS of Eq. (\ref{firstlaw}) with its internal energy $u_i$. Next, we note that the Brownian particle is driven far from equilibrium in two ways:
\begin{itemize}
\item[$(i)$] The harmonic trap follows a pre-fixed, time-dependent, time-periodic protocol that drives the particle away from equilibrium.  When the time-dependent trap does thermodynamic work on the Brownian particle, it is called {\it{work done on the system}} and is positive by our convention. When the opposite happens, i.e. the Brownian particle does work on the trap, {\it work is done by the system}. This is nothing but the work extracted from the system. Nevertheless, in either case the time-dependent change of the trap contributes to the thermodynamic work. If the trap does not change time-dependently (i.e, $\dot \kappa=0$), thermodynamic work becomes zero. This motivates us to identify the second term on RHS of Eq. (\ref{firstlaw}) as the work $w_i$. 
\item[$(ii)$] Together with the passive (medium) particles, the bath also contains active particles. Due to the collisions between the active particles and the working substance (i.e. the Brownian particle), the dynamics of the Brownian particle includes athermal noise that drives the particle away from the equilibrium, breaking the fluctuation dissipation relation between the viscous drag and the noise strength. Nevertheless, the collisions are random and we assume that it contributes to the heat, similar to the thermal collisions by the passive (medium) particles.  Hence, a part of the heat $q_i$ contains contribution from the thermal fluctuations and the other part contains contribution from active/athermal fluctuations. 
\end{itemize}
It is important to mention here that the aforementioned identification is not unique and there are other possible identifications. For example in \cite{Cates22, Cates21EPL, Fodor20}, active degrees of freedoms are excluded from the definitions of heat and are instead included in the thermodynamic work. These approaches can draw qualitatively different conclusions, such as, vanishing efficiency at large cycle time, which we will not obtain here. Instead, it will be shown later that in our case we obtain finite, non-zero, Carnot-bound efficiency at large cycle time.

Identification of heat in such non-equilibrium active micro heat engines is indeed a debated topic. In general, as shown in \cite{Steffenoni, Viktor20}, if one can map a non-equilibrium cyclic active heat engine to an effective equilibrium description then one can clearly identify the energy current in the system as thermodynamic heat. In such cases one may have very high effective temperatures (also observed in recent experiment \cite{Krishnamurty16}), that boost the efficiency of the active engines beyond their passive counterparts, but still bounded by the second law. However, if such mapping does not exist then the efficiency of these engines can break the second law bound \cite{Fodor20}, and are termed as work-to-work converters \cite{Horowitz16}.

It may also be noted here that this simple model does not allow us to consider  the energy spent on the active particles for maintaining their activity directly. Instead, later we will see that it is indirectly taken care of by the finite correlation time of the active noise $\eta(t)$.  We believe that the energy required to maintain activity can be taken care of directly by considering the so called depot models of active Brownian particles and is beyond the scope of the present work \cite{DepotMod}. 

For our model the average work is given as

\beqa
W_i=\langle w_i(t)\rangle=\frac{1}{2}\int dt'\dot \kappa(t')\langle x_i^2(t')\rangle=\frac{\gamma_i}{2}\int dt'\left(\frac{\ddot f_i(t')}{f_i(t')}-\frac{\dot f_i^2(t')}{f_i^2(t')}\right)\sigma^{(i)}(t'),
\label{work}
\eeqa    
where $\sigma^{(i)}(t)\equiv \langle x_i^2(t)\rangle$, where $i=1, 2$ corresponds to two halves of the cycle as mentioned above. Similarly the average heat dissipated is
\beqa
Q_i=\langle q_i(t)\rangle = -\frac{1}{2}\int dt' \kappa(t')\dot\sigma^{(i)}(t')=-\frac{\gamma_i}{2}\int dt'\left(\frac{\dot f_i(t')}{f_i(t')}\right)\dot\sigma^{(i)}(t'),
\label{heat}
\eeqa
and the average internal energy is
\beqa
\langle U_i(t)\rangle = \frac{1}{2}\kappa(t)\sigma^{(i)}(t).
\label{inten}
\eeqa
Also the average efficiency, power and rate of average entropy production are respectively defined as 
\beqa
\eta = -\frac{(W_1~+~W_2)}{Q_1}, 
~~P = \frac{W_1~+~W_2}{\tau}, 
~~\dot S =-\sum_{i=1}^2\frac{1}{T_i}\frac{dQ_i}{dt}= \frac{1}{2}\sum_{i=1}^2\frac{\gamma_i}{T_i}\left(\frac{\dot f_i(t)}{f_i(t)}\right)\dot\sigma^{(i)}(t), \nonumber\\
\label{eff}
\eeqa 
An important thermodynamic parameter is efficiency at maximum power output where one can calculate efficiency $\eta$ at a particular cycle time for which $P$ becomes maximum. Therefore, to calculate average efficiency at maximum power, first we can find a $\tau=\tau^*$ that satisfies $\frac{\partial P}{\partial\tau}=0$. 
All the average thermodynamic quantities involve $\sigma^{(i)}$ that can be calculated from Eq. (\ref{formal1}) as
\beqa
\sigma^{(i)}(t)&=& \frac{\langle x_{i}(0)^2\rangle}{f_i^{2}(t)} + \frac{2k_BT_i}{\gamma_if_i^{2}(t)}\int^t dt' f_i^{2}(t') + \frac{1}{\gamma_i^2f_i^{2}(t)}\int^t \int^t dt' dt''\langle \eta_i(t')\eta_i(t'')\rangle~f_i(t')f_i(t''),\nonumber\\
\label{sigma}
\eeqa 
where we have used the noise correlations given in Eq. (\ref{noise}). For a given $f(t)$, now we can calculate $\sigma^{(i)}(t)$ and thereby all the average thermodynamic quantities exactly. This is not in general possible in case of similar models due the non-linear nature and explicit time dependence of the trap strength $\kappa(t)$ in the problem. 

\section{The protocol}
In this section we discuss the class of protocols that can be studied. We also motivate a particular protocol that we use in our analysis. From Eq. (\ref{protocol1}), it is clear that any protocol of type
\begin{equation}
    \kappa(t) = \gamma_i\frac{\dot{f_i}(t)}{f_i(t)},
    \label{protoform}
\end{equation}    
forms a class of protocols that in principle when substituted in Eq. (\ref{sigma}) should provide the exact expressions for $\sigma^{(i)}(t)$ and other thermodynamic quantities for arbitrary cycle time $\tau$. 
For the model to work as an engine only condition is that the protocol $\kappa(t)$ has to be a decreasing function of time in the first half and an increasing function of time in the second half of the cycle (and continuous at $t=\tau/2$). The simplest protocol that satisfies these conditions is of type $1/t$ as shown in Fig. \ref{Fig.1} and given by Eq. (\ref{Eq.14}) \cite{Engel09}. We also believe such a protocol can also be realized in an experiment \cite{Bechinger12, Krishnamurty16}. Effect of more complicated protocols on the thermodynamic entities mainly on the efficiency may also be interesting and will be studied elsewhere. Thus, we consider
\begin{equation}
\kappa(t)= 
\begin{cases}
\frac{k_{1}k_{2}}{k_{2}+(k_{1}-k_{2})\frac{2t}{\tau}} & , 0 \leq t \leq \tau/2,\\
\frac{k_{1}k_{2}}{k_{1}+(k_{1}-k_{2})(1-\frac{2t}{\tau})} & , \tau/2< t \leq\tau,
\end{cases}
\label{Eq.14}
\end{equation}
where, $k_{1}$ and $k_{2}$ are the maximum and minimum value of the trap strength and $\tau$ is the total cycle time. The trap strength decreases in the first half, which in the presence of passive thermal bath, mimics the isothermal expansion of the system. Similarly in the second half it increases mimicking the isothermal compression of the system. At $\tau/2$ the temperature suddenly drops from $T_{1}$ to $ T_{2}$ whereas at $\tau$ it suddenly jumps from $T_2$ to $T_1$. The expansion and compression steps connected by sudden temperature jumps are known as Stirling-type protocol for micro-heat engine \cite{Bechinger12, Arnab19}. Using $\kappa(t)$ we obtain $f_i(t)$ as 
\begin{equation}
f(t)=\begin{cases}
f_1(t) =\left(\frac{[k_{2}+(k_{1}-k_{2})\frac{2t}{\tau}]}{k_{2}}\right) ^{a_1}  & , 0\leq t\leq\tau/2,\\ f_2(t)= \left(\frac{[k_{1}+(k_{1}-k_{2})(1-\frac{2t}{\tau})]}{k_{1}}\right) ^{-a_2} & , \tau/2 \leq t \leq \tau,
\end{cases}
\end{equation}
where $a_i=\frac{\tau k_{1}k_{2}}{2\gamma_i(k_{1}-k_{2})}$. Next, using the expressions of $f_i(t)$ we will calculate the relevant thermodynamic quantities of passive as well as active engines, as defined in the previous section.
\section{Passive Micro Heat Engine}

\subsubsection{Time evolution of the Variance of the particle in presence of thermal fluctuation}
Here we consider the reservoirs to be passive (i.e. the active noise $\eta(t)=0$) and equilibrated at $T_{1, 2}$ in two halves of the cycle respectively. We will determine the expression for the variance in the position of the particle as a function of time. At time $t=0$, the particle starts from a given initial position $x(0)=0$. We  assumed that the process starts from equilibrium at time $t=0$ and then the particle is driven out of equilibrium by the time-periodic protocol. After several cycles of the protocol the system will reach the non equilibrium steady state (NESS). In NESS, for the first half of the cycle the expression of the variance is
\begin{eqnarray}
	\sigma_{Th}^{(1)}(t)&=&\frac{\sigma_{Th}(\tau)}{f_1(t)^{2}}
	+\frac{k_{B}T_{1}k_{2}\tau}{\gamma_1(k_{1}-k_{2})(2a_1+1)}\left(\frac{f_1(t)^{\frac{(1+2a_1)}{a_1}}-1}{f_1(t)^{2}}\right).
	\label{Eq.22}
\end{eqnarray}
Similarly the expression of variance in second half of the cycle becomes
\begin{eqnarray}
	\sigma_{Th}^{(2)}(t)&=&\frac{\sigma_{Th}(\frac{\tau}{2})}{f_{2}(t)^{2}}- \frac{k_{B}T_2k_{1}\tau}{\gamma_{2}(k_{1}-k_{2})(1-2a_{2})}\left(\frac{f_2(t)^{-\frac{(2a_2-1)}{a_2}}-1}{f_2(t)^{2}}\right). \nonumber\\
	\label{Eq.23}
\end{eqnarray}
The subscript $Th$ corresponds to the {\it thermal} fluctuations. Also, 
\begin{eqnarray}
\sigma_{Th}(\tau)&=&\frac{\left[\frac{k_{B}T_{1}\tau k_{1}}{\gamma_{1}(k_{1}-k_{2})(2a_{1}+1)}\left(\frac{k_{2}}{k_{1}}\right)^{2a_{2}}\left(1-\left(\frac{k_{2}}{k_{1}}\right)^{2a_{1}+1}\right)+\frac{k_{B}T_{2}\tau k_{2}}{\gamma_{2}(k_{1}-k_{2})(2a_{2}-1)}\left(1-\left(\frac{k_{2}}{k_{1}}\right)^{2a_{2}-1}\right)\right]}{\left[1-\left(\frac{k_{2}}{k_{1}}\right)^{2(a_{1}+a_{2})}\right]},\nonumber \\
\sigma_{Th}(\tau/2)&=&\left(\frac{k_{2}}{k_{1}}\right)^{2a_{1}}\sigma_{Th}(\tau)+\left[\frac{k_{B}T_{1}\tau k_{1}}{\gamma_{1}(k_{1}-k_{2})(2a_{1}+1)}\left(1-\left(\frac{k_{2}}{k_{1}}\right)^{2a_{1}+1}\right)\right],    
\end{eqnarray}
are calculated by the continuity of $\sigma_{Th}^{(i)}$'s at $t=\tau/2$ and $t=\tau$. We also simulate this dynamics using a standard Langevin integrator for comparison with analytical expressions of the variance. This fixes the benchmark of our numerical calculations as well. Below we briefly describe all the simulation results and match them with analytical expressions. We take $k_{B}=1$.
\begin{figure}[!t]
\includegraphics[width=8cm,height=6.5cm,angle=0]{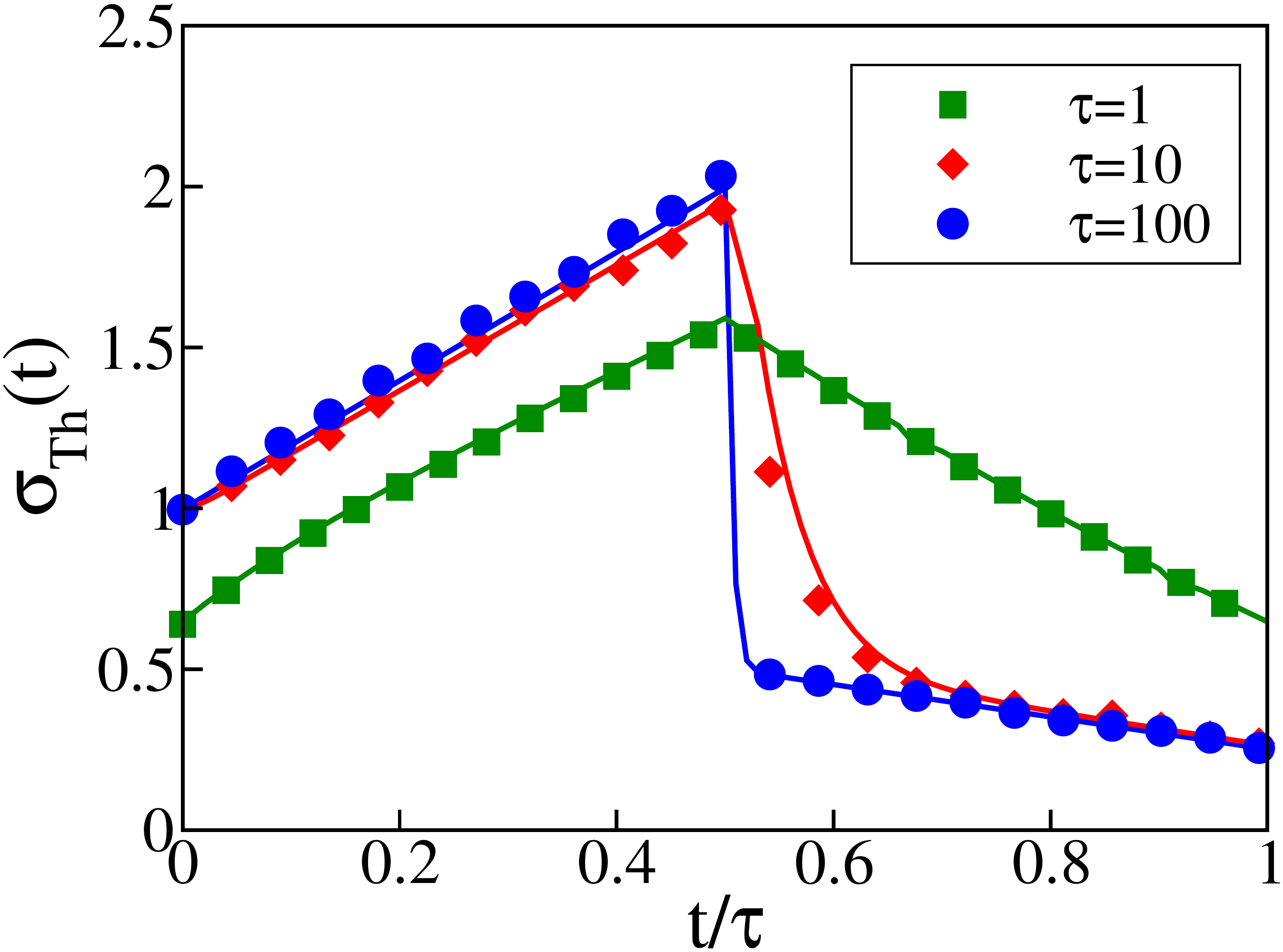}
    \hspace{0.2cm}
\includegraphics[width=8cm,height=6.5cm,angle=0]{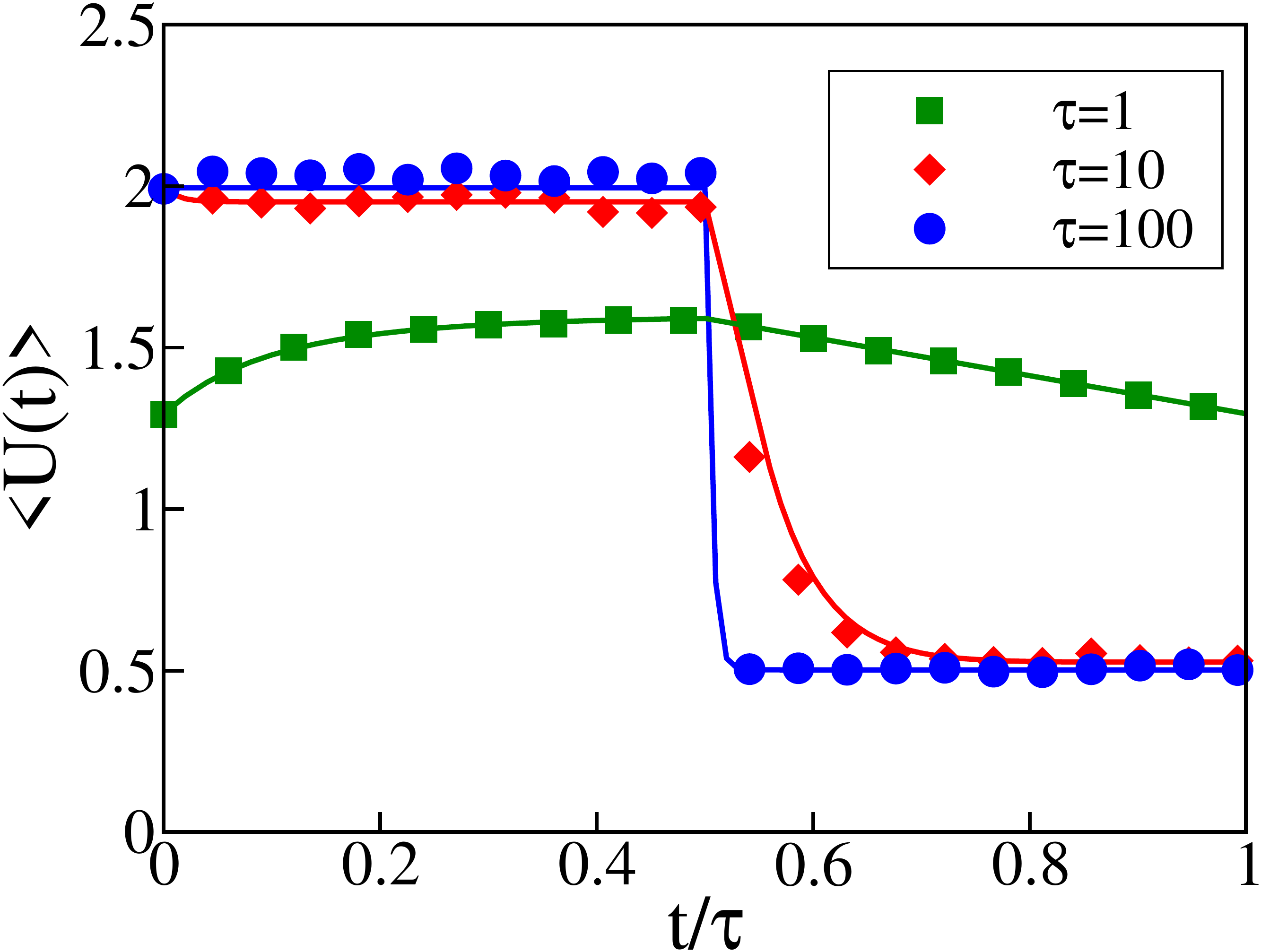}
\caption{Left panel represents the mean squared displacement $\sigma_{Th}(t)$ as a function of $\frac{t}{\tau}$.
  Right panel represents the internal energy of the particle $\langle U(t) \rangle$ versus $t/\tau$, for three cycle times $\tau=1, 10, 100$. The parameters used are $k_{1}=4.0$, $k_{2}=2.0$, $\gamma_{1}=1.0$, $\gamma_{2}=2.0$, $T_{1}=4.0$, $T_{2}=1.0$. In all the figures, unless otherwise specified, symbols represent simulation results and solid lines analytical or numerical results.} 
\label{Fig.2}  
\end{figure}
In Fig. \ref{Fig.2} we plot the mean squared displacement and internal energy of the particle as a function of the time for different values of the cycle time $\tau$. In the first half of the cycle, during the isothermal expansion, due to the decreasing stiffness of the trap, the accessible volume for the particle increases. This leads to the increase in the variance $\sigma_{Th}$ of the position of the particle. Similarly during the isothermal compression in the second half, the trap strength increases which reduces the accessible volume for the colloid. Hence the variance $\sigma_{Th}$ decreases. We also plot the internal energy $\langle U_{Th}(t)\rangle=\frac{1}{2} \kappa(t)\sigma_{Th}(t)$, which saturates to a constant value that is equal to $T_1/2$, in the first half and to $T_2/2$ in the second half satisfying equi-partition theorem for larger values of the cycle time $\tau$ implying quasistatic process. However, for smaller values of $\tau$ (non-quasistatic case) there is no equipartition and system is far away from equilibrium. We note that $\Delta \sigma_{Th}(\tau/2)=\sigma_{Th}(\frac{\tau}{2}^{+})-\sigma_{Th}(\frac{\tau}{2}^{-})$  and similarly $\Delta \sigma_{Th}(\tau)$ increase with $\tau$. The numerical results match with the analytical expression extremely well. In the Fig. \ref{Fig.50}, we  plot $\sigma_{Th}(t)$ with the protocol $\kappa(t)$ for different cycle times $\tau$. One can see that in the right panel of Fig. \ref{Fig.50}, it mimics the pressure-volume diagram of a macroscopic Stirling engine. In the left panel of this diagram we have plotted $\sigma_{Th}(t)$ versus $\kappa(t)$ with small values of $\tau$ (non-quasistatic case). It is evident from the plots that the jumps of $\sigma_{Th}$ at $\tau$ and $\tau/2$ reduce as we reduce $\tau$.
\begin{figure}[t]
    \hspace{-1cm}
    \includegraphics[width=8cm,height=6.5cm,angle=0]{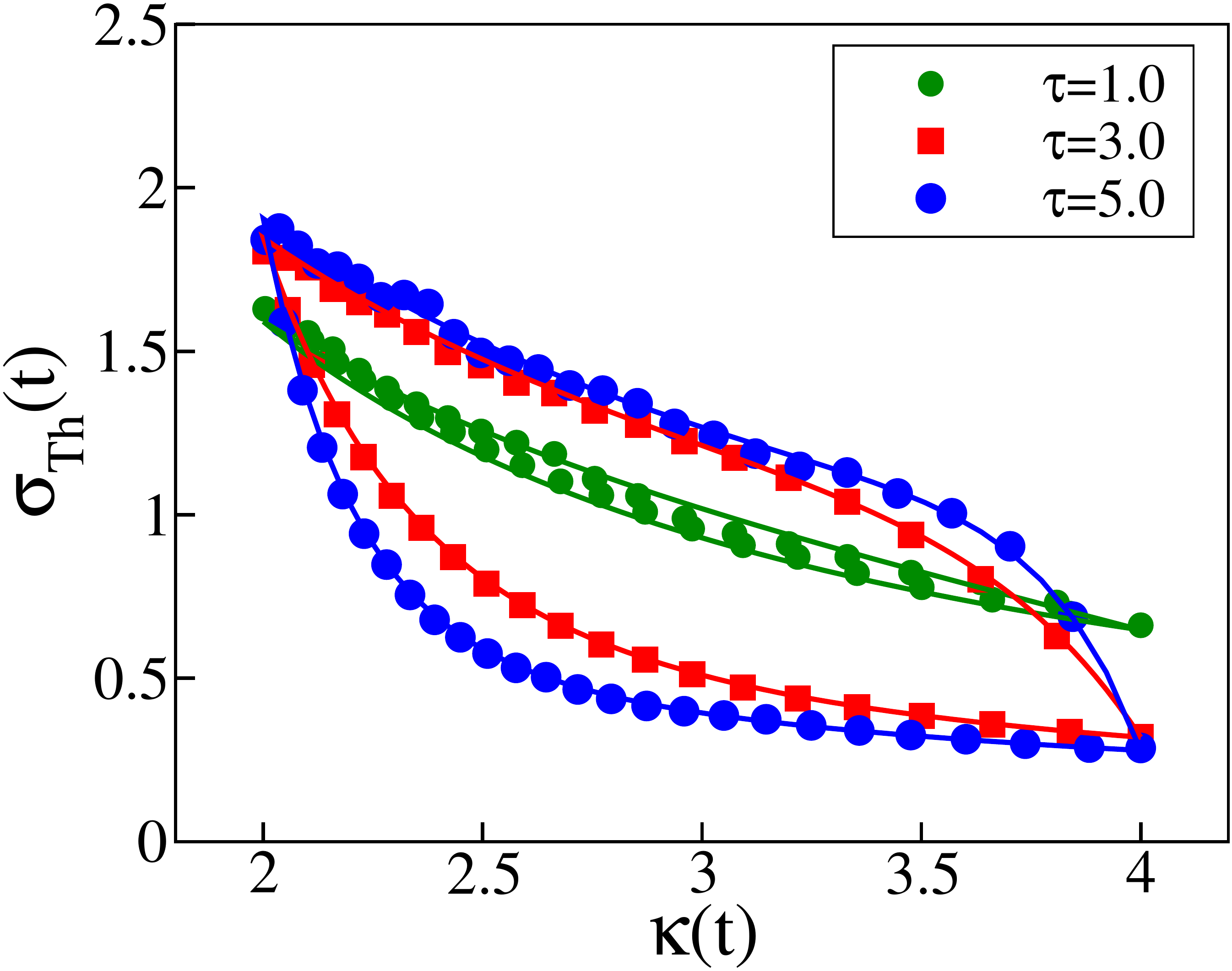}
    \includegraphics[width=8cm,height=6.5cm,angle=0]{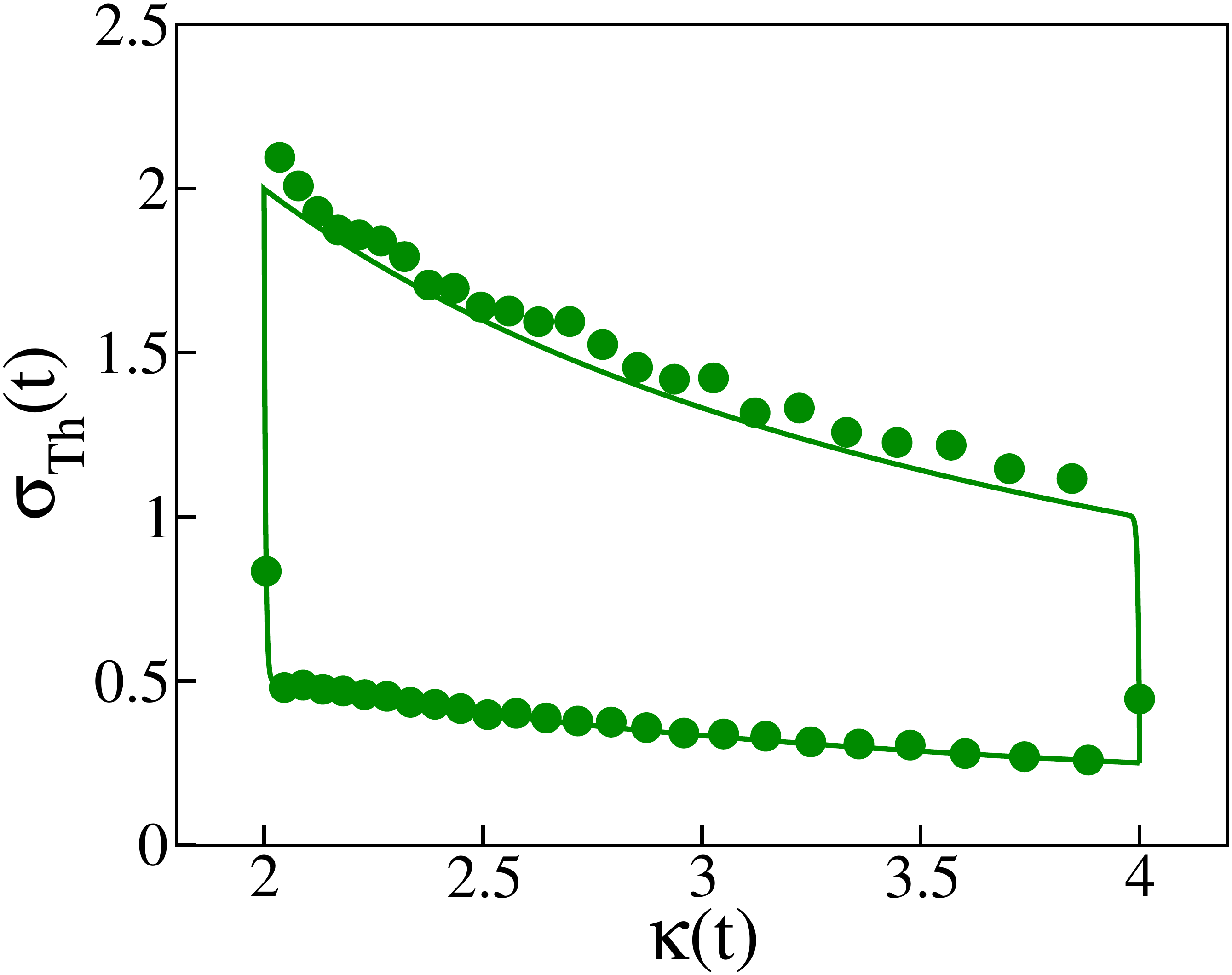}
    \caption{Plot of $\sigma_{Th}(t)$ versus $\kappa(t)$. Left panel for 
      different $\tau$ for non-quasi static cases. Right panel for large $\tau$, the quasistatic limit. The parameters used are $k_{1}=4.0$, $ k_{2}=2.0$, $\gamma_{1}=1.0$, $\gamma_{2}=2.0$, $T_{1}=4.0$, $T_{2}=1.0$. In the right panel $\tau=300$.}
\label{Fig.50}  
\end{figure}
\subsubsection{Analytical calculation of the Thermodynamic quantities}
Next we calculate average thermodynamic work and heat from their definitions, using the expression of $\sigma_{Th}^{(i)}(t)$ for both the halves of a cycle. Using these we will also be able to calculate the efficiency. It is to be noted here that through out the paper we have derived the expressions for all the noise-averaged thermodynamic quantities (work, heat, entropy etc.) as accumulated up to time $t \leq \tau$ by integrating Eq. (\ref{work}, \ref{heat}, \ref{eff}) accordingly. These analytical expressions for passive as well as active cases are provided below in the corresponding sections of the paper, where putting $t=\tau$ or $t=\tau/2$, we estimate the accumulated value of a noise-averaged thermodynamic quantity ( i.e. either work, heat or entropy) for a full or half cycle respectively. From the definition of work in Eq. (\ref{work}) and from Eqs. (\ref{Eq.22}, \ref{Eq.23}) we get 
\begin{eqnarray}
W_{Th}^{(1)}(t)&=&\frac{k_{1}\left(f_{1}(t)^{-\frac{(2a_{1}+1)}{a_{1}}}-1\right)}{2(2a_{1}+1)}\left[\sigma_{Th}(\tau)-\frac{k_{B}T_{1}k_{2}\tau}{\gamma_{1}(k_{1}-k_{2})(2a_{1}+1)}\right]~-~\frac{k_{B}T_{1}}{(2a_{1}+1)}\log\left[f_{1}(t)\right]\nonumber\\
	\label{Eq.25}
\end{eqnarray}
and
\begin{eqnarray}
W_{Th}^{(2)}(t)&=&\frac{k_{2}\left(f_{2}(t)^{-\frac{(1-2a_{2})}{a_{2}}}-1\right)}{2(1-2a_{2})}\left[\sigma_{Th}(\frac{\tau}{2})-\frac{k_{B}T_{2}k_{1}\tau}{\gamma_{2}(k_{1}-k_{2})(1-2a_{2})}\right]~-~\frac{k_{B}T_{2}}{(1-2a_{2})}\log\left[f_{2}(t)\right]\nonumber\\
	\label{Eq.26}
\end{eqnarray}
At large cycle time $\tau \rightarrow\infty$ the work done in the first half of the cycle is  $W_{Th}^{(1)}=-\frac{k_{B}T_{1}}{2}\log[\frac{k_{1}}{k_{2}}] $, and the $2^{nd}$ half is $W_{Th}^{(2)}=\frac{k_{B}T_2}{2}\log[\frac{k_{1}}{k_{2}}] $. So the total thermodynamic work along a cycle in quasistatic limit is, $W_{total}=\frac{k_{B}}{2}(T_{2}-T_{1})\log{\frac{k_{1}}{k_{2}}}$. The internal energy of the particle defined as $\langle U(t)\rangle=\frac{1}{2}\kappa(t) \sigma_{Th}^{(i)}(t)$ , the average heat dissipation as defined in Eq. (\ref{heat}) and using the first law of stochastic thermodynamics, we calculate average heat along the first half of a cycle as
\begin{eqnarray}
	&&Q_{Th}^{(1)}=\frac{1}{2}\left[\kappa(\frac{\tau}{2})\sigma^{(1)}_{Th}(\frac{\tau}{2})-\kappa(0)\sigma^{(1)}_{Th}(0)\right]-W^{(1)}_{Th}(\frac{\tau}{2})
	\label{Eq.32}
\end{eqnarray}
Similarly, during the contraction step we get
\begin{eqnarray}
	Q_{Th}^{(2)}&=&\frac{1}{2}\left[\kappa(\tau)\sigma^{(1)}_{Th}(\tau)-\kappa(\frac{\tau}{2})\sigma^{(2)}_{Th}(\frac{\tau}{2})\right]-W^{(2)}_{Th}(\tau)
	\label{Eq.33}
\end{eqnarray}
\begin{figure}[!t]
  \hspace{-1cm}
\includegraphics[width=0.5\textwidth]{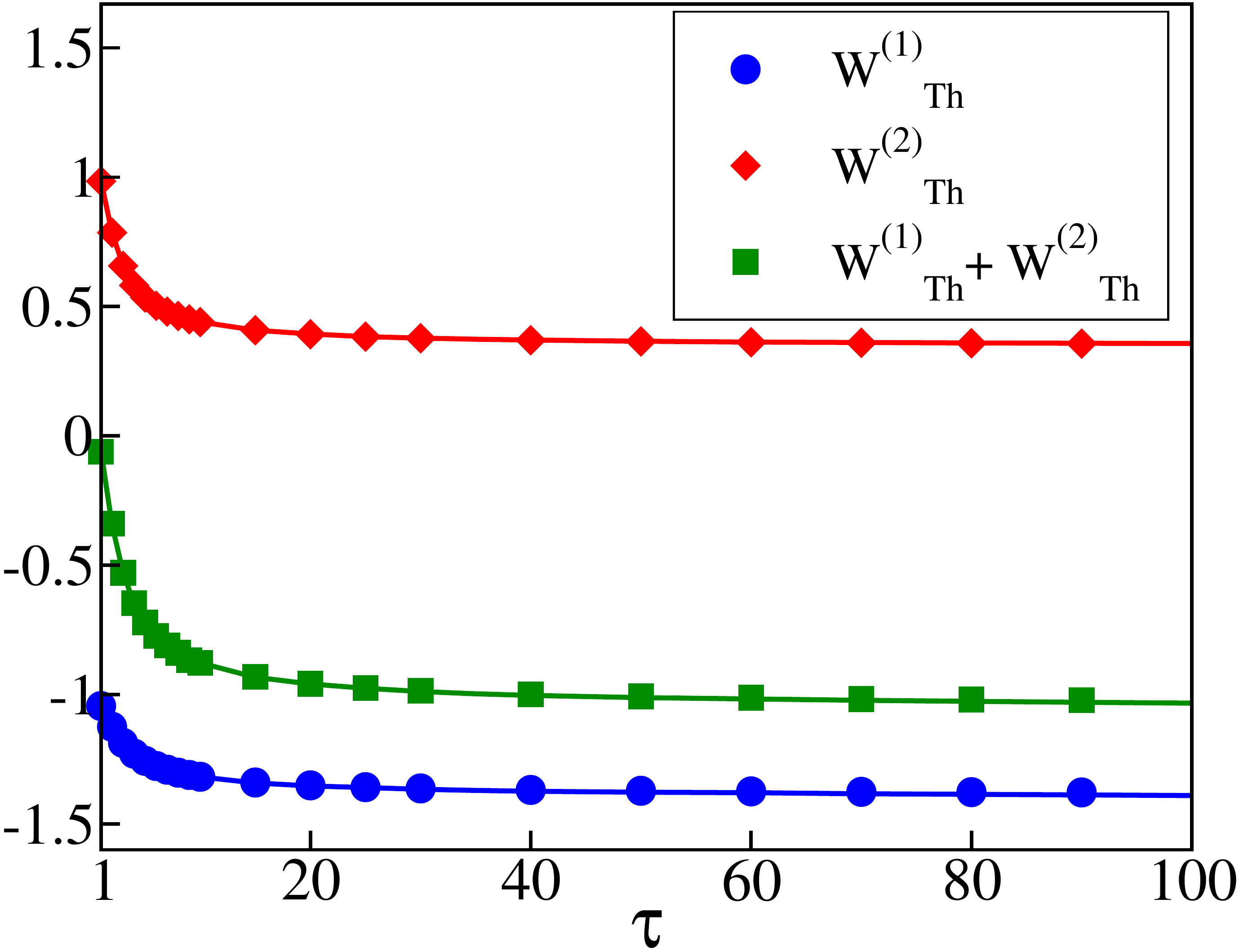}
\includegraphics[width=0.5\textwidth]{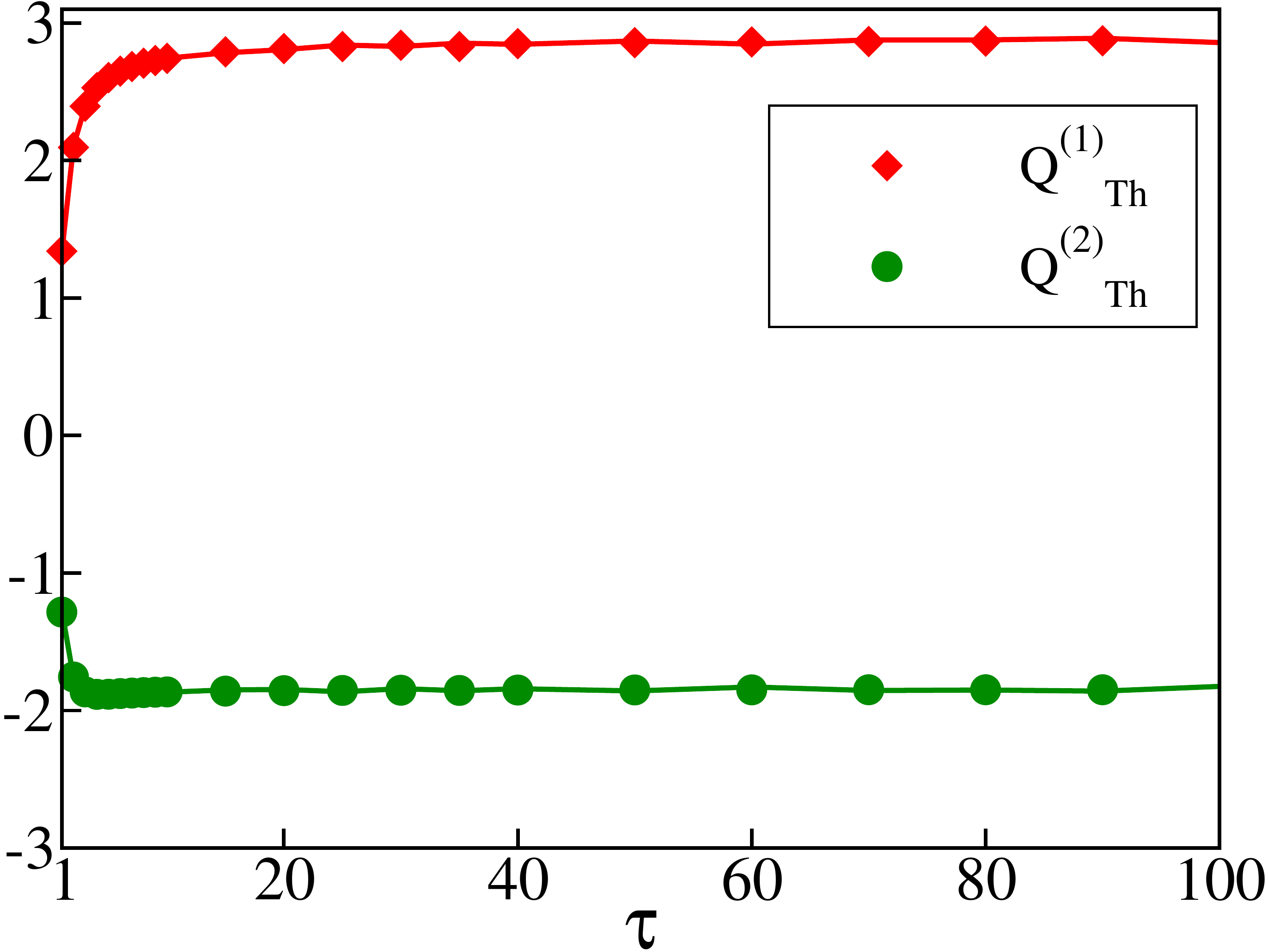}
\caption{Individual works and total work versus the cycle time $\tau$ (left panel). Heat absorbed,  dissipated $Q_{Th}^{(1)}$ and $ Q_{Th}^{(2)}$ respectively as a function of cycle time $\tau$ (right panel). The parameters are $k_{1}=4.0$, $ k_{2}=2.0$, $\gamma_{1}=1.0$, $\gamma_{2}=2.0$, $T_{1}=4.0$, $T_{2}=1.0$.}
\label{Fig.4}
\end{figure}

In Fig. \ref{Fig.4}, we have plotted average accumulated work and heat within a cycle, as a function of $\tau$, obtained from analytics as well as from simulation.
It is clear form the graph that the total work $W_{Th}^{(1)} + W_{Th}^{(2)}$ initially decreases and then saturates at a certain negative value (quasistatic value), suggesting that the model indeed works as an engine. Similarly in Fig. \ref{Fig.4} (right panel), we plot heat accumulated within a cycle time $\tau$, as a function of $\tau$, obtained both from analytics and from simulation. $Q_{Th}^{(1)}$ is the heat absorbed by the system from the hot bath. $Q_{Th}^{(2)}$ is the heat released in the cold bath. As a function of $\tau$, $Q_{Th}^{(1)}$ initially increases and then saturates at some positive value and similarly, $Q_{Th}^{(2)}$ initially decreases and saturates at a negative value.

\begin{figure}[!t]
    \hspace{-1cm}
    \includegraphics[width=0.5\textwidth, height=6.5cm]{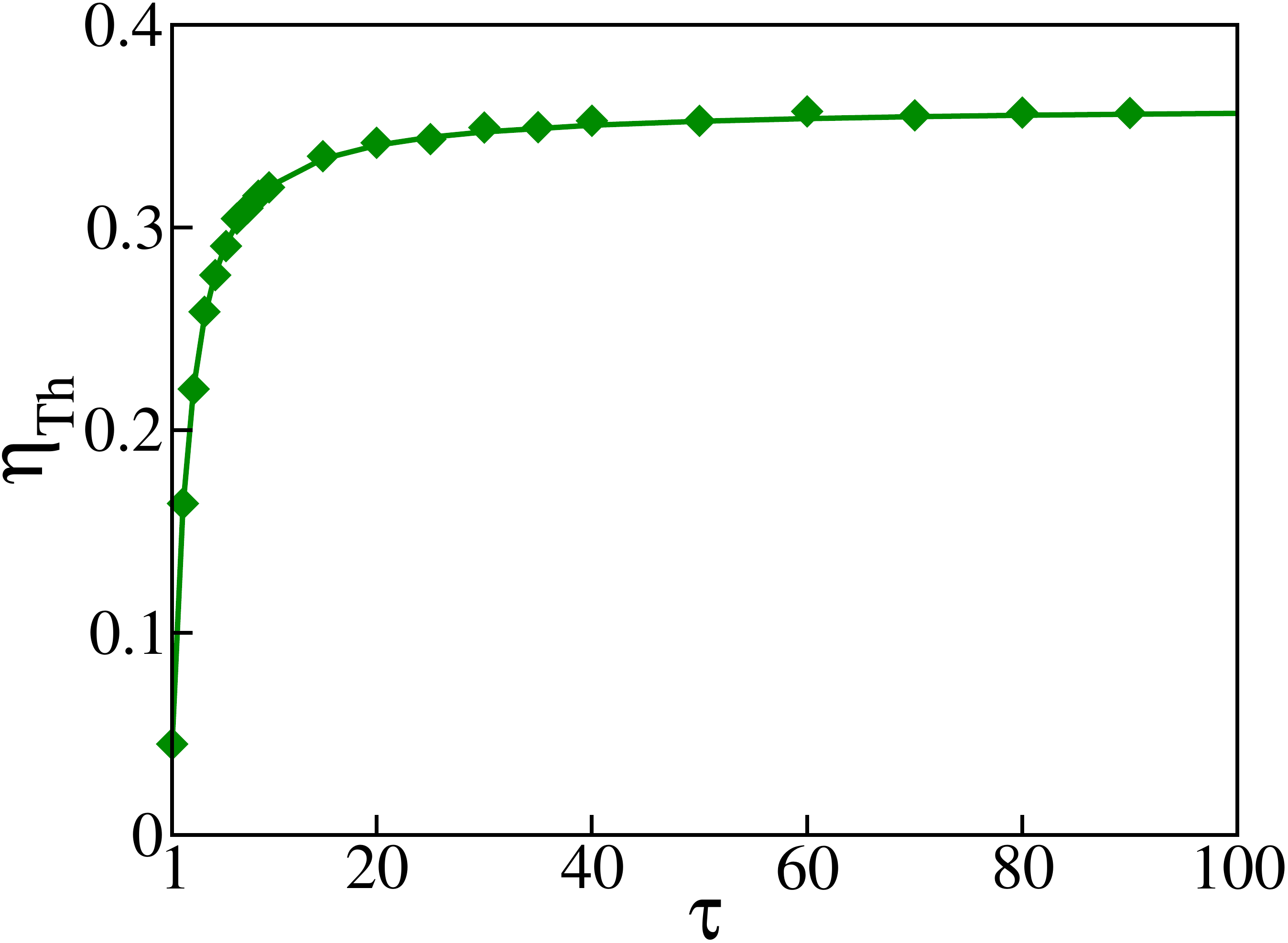}
    \includegraphics[width=0.5\textwidth, height=6.5cm]{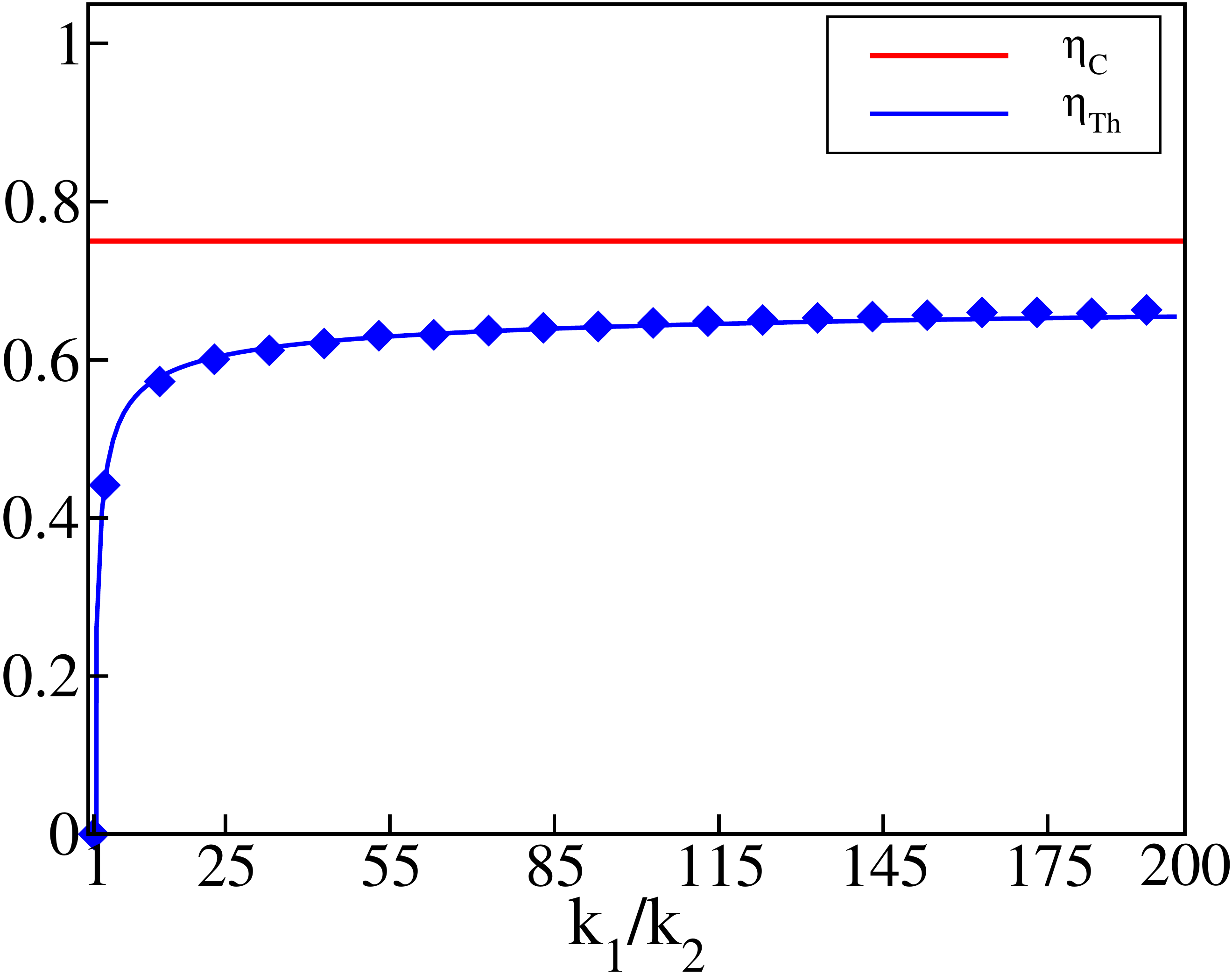}
    \caption{The efficiency $\eta_{Th}$ as a function of cycle time $\tau$ (left panel). Efficiency in comparison with the Carnot efficiency as a function of the ratio of maximum and minimum value of the trap strength $\frac{k_1}{k_2}$ in the quasistatic case (right panel). Interestingly the efficiency reaches the Carnot efficiency for large values of the ratio. Parameters used are $k_{1}=4.0$, $k_{2}=2.0$, $\gamma_{1}=1.0$, $\gamma_{2}=2.0$, $T_{1}=4.0$, $T_{2}=1.0$ (left panel) and in the right panel $k_{2}=2.0$ is fixed and $k_1$ varied for $\tau=200$.}
\label{effic}  
\end{figure}

From noise-averaged total work and absorbed heat, it is straight forward to calculate the efficiency of the engine, defined as : $\eta_{Th}(t)=-\frac{W_{Th}^{(1)}(t)+W_{Th}^{(2)}(t)}{Q_{Th}^{(1)}(t)}$. In Fig. \ref{effic}(left panel), we plot efficiency as a function of the cycle time. In the quasi-static limit ($\tau\rightarrow \infty$) with straightforward algebra one can show that the expression for efficiency can be reduced to
\begin{eqnarray}
  \eta_{Th}= \frac{\frac{k_{B}}{2}(T_{1}-T_{2})~\log{\frac{k_{1}}{k_{2}}}}{\frac{k_{B}}{2}(T_{1}-T_{2})+\frac{k_{B}T_{1}}{2}\log{\frac{k_{1}}{k_{2}}}}
=\frac{\eta_{C}}{1+\frac{\eta_{C}}{\log{\frac{k_{1}}{k_{2}}}}}.
\label{Eq.34}
\end{eqnarray}
The quasistatic efficiency of the system is plotted in Fig. \ref{effic} (right panel), with the ratio of the maximum and minimum values of the protocol. It is evident from Fig. \ref{effic} that the engine performs poorly when compared to the Carnot efficiency even in the quasistatic limit. However, our engine reaches Carnot efficiency as the ratio of maximum and minimum value of the trap strength approaches infinity logarithmically. This indicates that the performance of the engine depends on the space available for the particle to move around \cite{Krishnamurty16}.
\begin{figure}[!t]
      \hspace{-1cm}
  \includegraphics[width=0.5\textwidth, height=6.5cm]{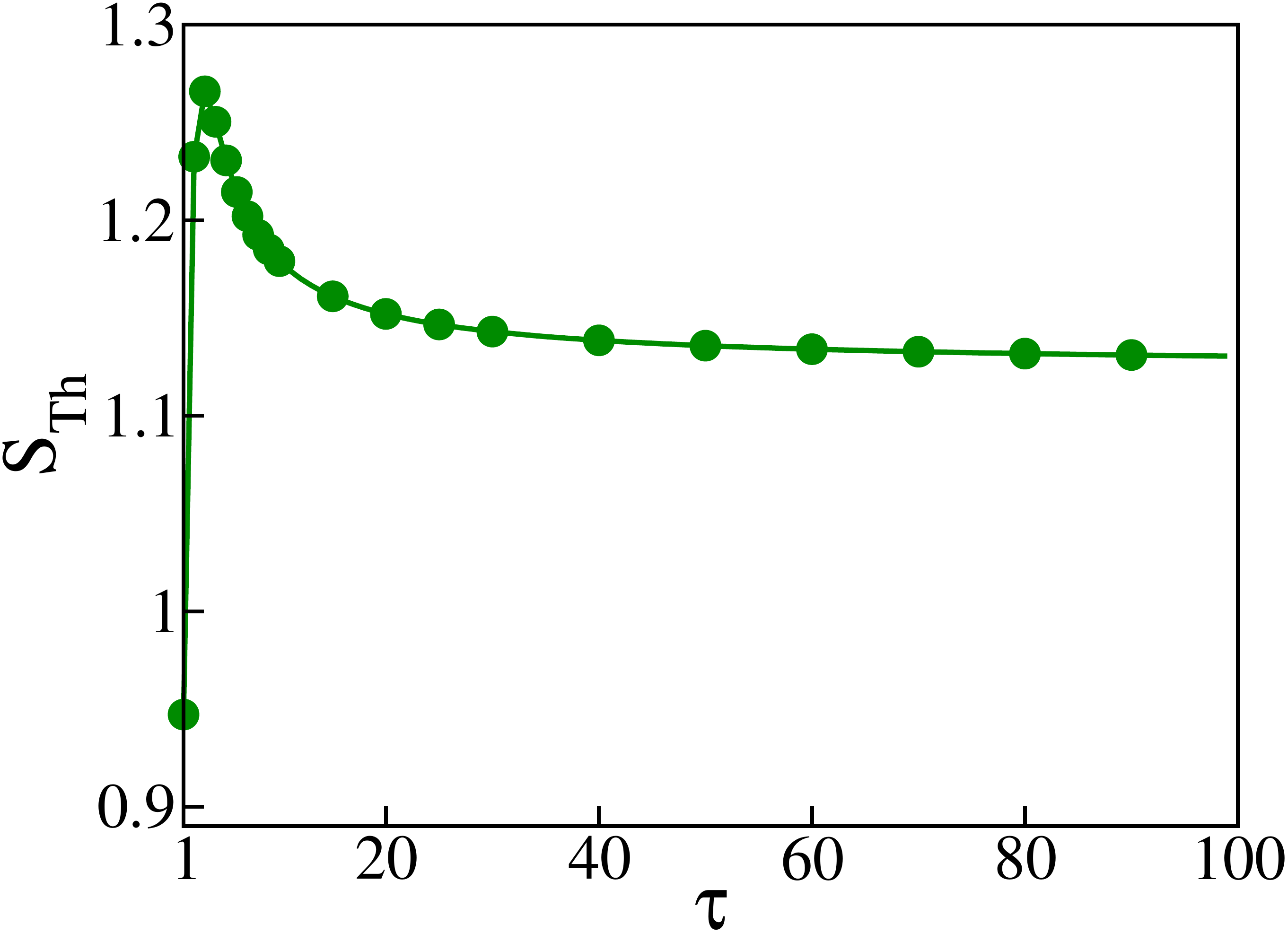}
  \includegraphics[width=0.5\textwidth, height=6.5cm]{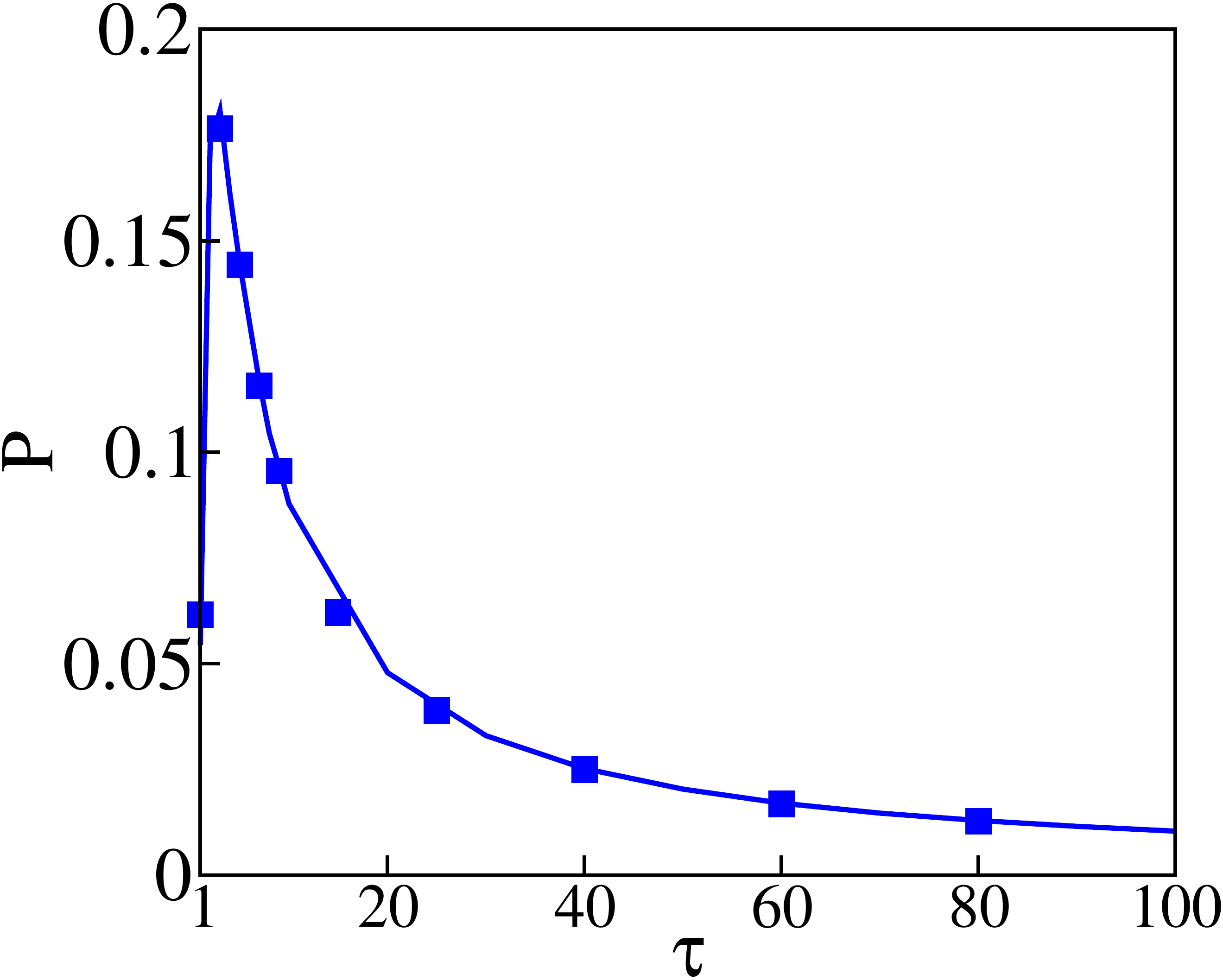}
\caption{Entropy (left panel) and power (right panel) as a function of the cycle time $\tau$. Parameters used are $k_{1}=4.0$, $ k_{2}=2.0$, $\gamma_{1}=1.0$, $\gamma_{2}=2.0$, $T_{1}=4.0$, $T_{2}=1.0$.}
   \label{entr}
\end{figure}
Next we calculate the entropy. The entropy production rate can be defined as
\begin{eqnarray}
\dot{S}^{(i)}_{Th}(t)= -\frac{1}{2T_{i}} \kappa(t)\dot{\sigma}^{(i)}_{Th}(t).
\label{entropyTh}
\end{eqnarray}
The expression for $\dot{\sigma}_{Th}^{(i)}$ can be obtained by multiplying Eq. (\ref{eom}) by $x_i$ on both sides. Then using it in Eq. (\ref{entropyTh}) and integrating it up to time $t$ we calculate entropy accumulated up to time $t$
\begin{eqnarray}
S^{(i)}_{Th}(t)&=& \frac{1}{T_{i}}\int^{t}\left(\frac{\kappa^{2}(t')\sigma^{(i)}_{Th}(t')}{\gamma_{i}}~-~\frac{\kappa(t')k_{B}T_{i}}{\gamma_{i}}\right)~dt',
\label{enTH}
\end{eqnarray}
where $i=1, 2$ correspond to different halves of the protocols and the total average entropy will be the sum of these two parts. It is plotted in Fig. \ref{entr}(left panel), along with the simulation results. Here we can see that the entropy is positive and reaches a maximum and saturates in the quasistatic case. We also plot power (defined in Eq. (\ref{eff})) as a function of the cycle time. We note that the power is maximum at certain value of $\tau$ where the $\eta_{Th}\sim 0.22$. It is much lower than the efficiency at quasistatic limit, indicating power-efficiency trade-off. In average power versus  average efficiency plot (Fig. \ref{PowerEfficiency}) the power-efficiency trade-off becomes apparent.  
\begin{figure}[!t]
      \hspace{-1cm}
  \includegraphics[width=0.5\textwidth, height=6.5cm]{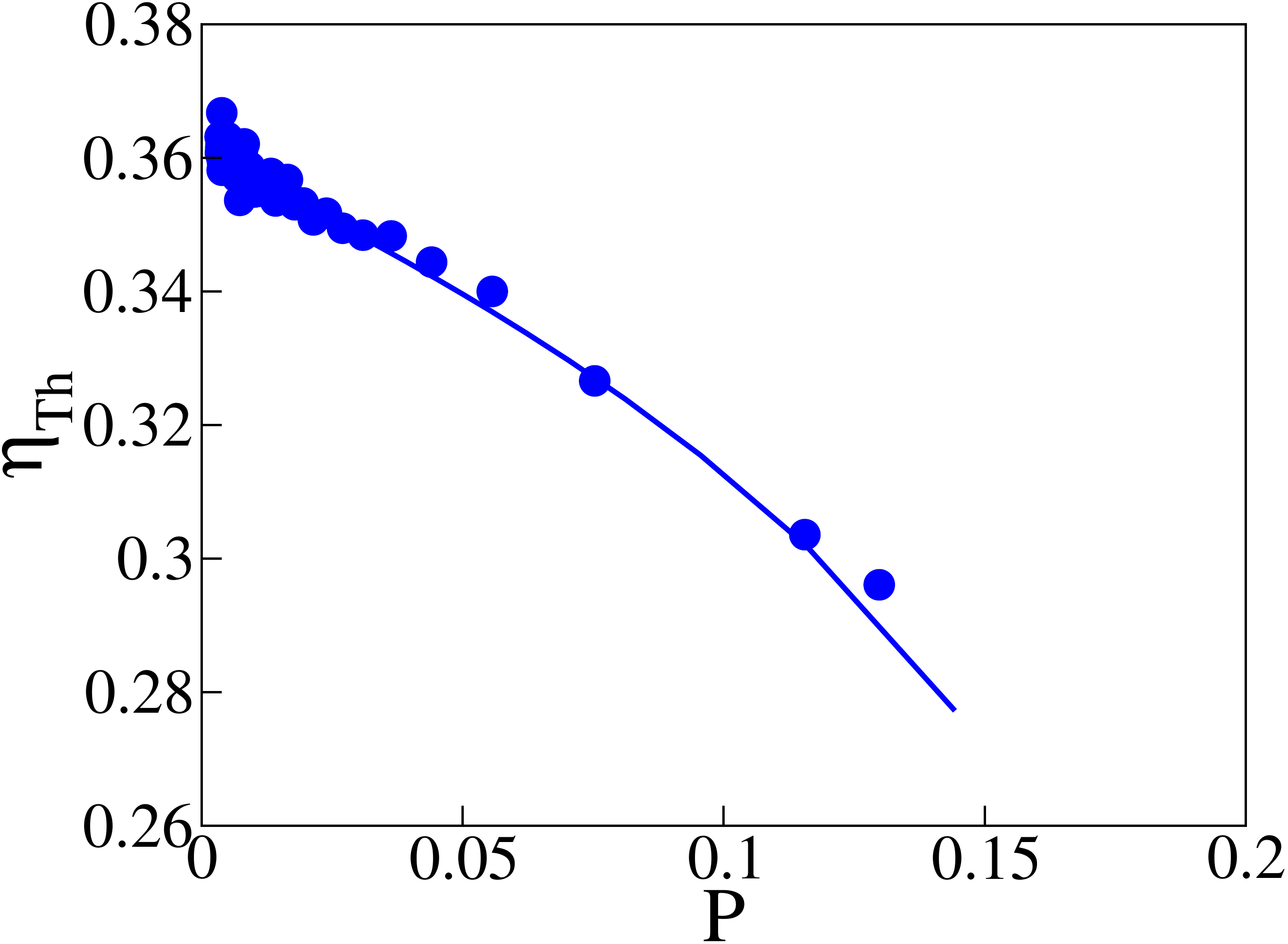}
 \caption{The efficiency ($\eta_{Th}$) vs power ($P$) for different values of $\gamma_{2}$. Parameters used are $T_{1}=4.0$, $T_{2}=1.0$, $k_{1}=4.0$, $ k_{2}=2.0$, $\gamma_{1}=1.0$. The plot shows a clear trade-off between the power and the efficiency as discussed in the text.}
\label{PowerEfficiency}
\end{figure}
So far, we have exactly calculated all the thermodynamic quantities of interests in case of passive engine for an arbitrary cycle time. In most of the models studied earlier, non-linearity and explicit time dependence of the protocols makes it impossible to perform such calculations \cite{Slahiri20, Arnab18, Arnab20, Arnab19, Gomez21, Steffenoni}, except some special cases and limits \cite{Marathe07, Viktor20}. The results in the passive case provide a handle to extend our calculations towards active particle heat engines. Later we will also compare the performance of active heat engines with their passive counterparts, which is still a debated subject \cite{Zakine17}. 

\section{Active Particle Heat Engines} \label{ActivEngs}
\begin{figure}[t]
  \hspace{-1cm}
   \includegraphics[width=0.5\textwidth, height=6.5cm]{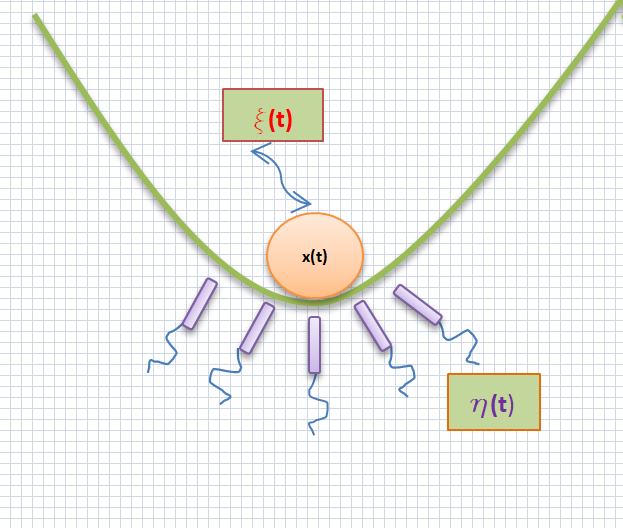}
  \includegraphics[width=0.5\textwidth, height=6.5cm]{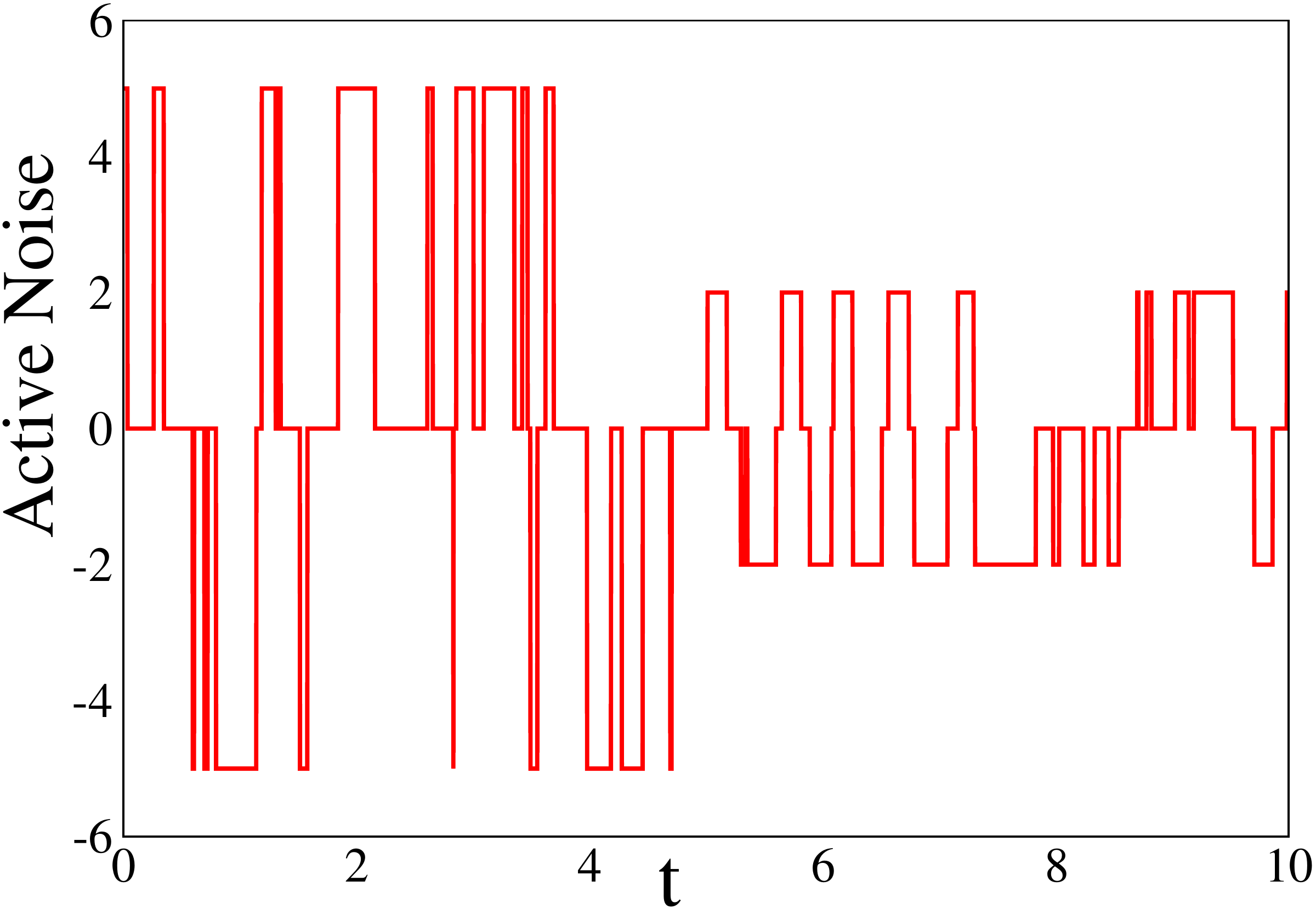}
    \caption{A schematic of a passive Brownian particle in an active non-equilibrium bath (left panel). A typical trajectory of trichotomous noise as a function of time $t$ representing the random force generated by the active particles. The noise strength can be larger in the first half than the second half of the cycle (right panel)}
\label{Fig.9}
\end{figure}

In this section we explore similar models of heat engines mentioned earlier but now with active fluctuations in addition to the thermal fluctuations. The set-up can be schematically shown in Fig. \ref{Fig.9}(left panel) where a passive colloidal particle gets kicked by the active entities like Bacteria or Janus particles suspended in the bath together with the thermal fluctuations. In the right panel of the same figure we plot a typical realization of the random force generated by the run and tumble like motion of the active particles.  Here we calculate the thermodynamics of active micro heat engine and compare their characteristics with passive micro-heat engine. This will also provide us with possible ways to enhance the performance of the active engines over the passive ones. In the following sections we have used 'RT' as a subscript of a quantity to imply 'Run and Tumble' type active fluctuations whereas 'CN' will be used to imply dynamics with active fluctuations represented by the Gaussian Coloured Noise. 
\subsubsection{Time evolution of the position variance of the particle in presence of RT-type active fluctuation}
Consider the situation depicted in Fig. \ref{Fig.9}(left panel), the trapped colloidal particle can get kicked from left or right due to the active run and tumble particles (during their run type of motion). We can think of this as the active particles are pushing the colloid in either direction for for random durations of time. There also might be some time intervals where the colloidal particle does not interact with active particles and only experiences thermal noise. Such a dynamics can be modeled by considering an active noise in Eq. (\ref{eom}) that can take three possible values randomly and switch among them at random time instances \cite{Chaki19, urna20}. In such a case of ``RT-type'' activity, the active noise $\eta_{i}(t)$ can choose three different values $A_i = (A_i^+, 0, A_i^-)$ randomly. We also take $A_i^-=-A_i^+$, so that the average active force acting on the particle will be zero. The values $A_i^{\pm}$ correspond to kicks from left and right respectively and $0$ implies no interaction of the colloid with the active particles. This noise is implemented in the following way in the simulations. Let at any time $t$, the value of $A_i$ be any of the $A_i^+$ or $A_i^-$, after a randomly chosen time interval $1/\lambda_i$, it goes to state $0$ with probability $1$, when it is in $0$ state, after another randomly chosen time interval it can go to $A_i^+$ or $A_i^-$ with probability $1/2$ each. The time interval for any of these flips is chosen from an exponential distribution. Here, $i=1, 2,$ again denotes the first and second half of the cycle and in general maximum value $A_i^{+}$ of $A_i$'s can be different in the two halves. It turns out that the noise-noise correlation in this case is exponential, and the strength of the correlation is proportional to the square of the maximum value of the noise strength ($A_i^{+}$), given by
\begin{eqnarray}
  \langle \eta_{RT}^{(i)}(t)\eta_{RT}^{(i)}(t')\rangle = \frac{\left(A_i^{+}\right)^2}{2} e^{-\lambda_i |t-t'|},
  \end{eqnarray}
where $\lambda_i$'s is the correlation rate. Even though the correlation is exponential the distribution of the noise is clearly non-Gaussian. 
In one dimension this corresponds to active particles undergoing RT dynamics that tumble between right-moving $(+)$ to left-moving $(-)$ states and vice-versa via a no-moving state with $A_i=0$. Here $A_i^{+/-}$ corresponds to the force imparted by the right/left moving active particle to the passive colloidal particle. We call this the {\it trichotomous noise}. A typical trajectory of the trichotomous noise is depicted in Fig. \ref{Fig.9}(right panel).

As in the passive case, the variance of the position of the colloid is important to calculate averaged thermodynamic quantities. We  calculate this from Eq. (\ref{formal1}, \ref{sigma}) with both the thermal and active noise. Other details of the model remain identical to the ones in the passive engine case. We can now calculate the variance as
\begin{figure}[t]
 \hspace{-1cm}
    \includegraphics[width=8.0cm,height=6.5cm,angle=0]{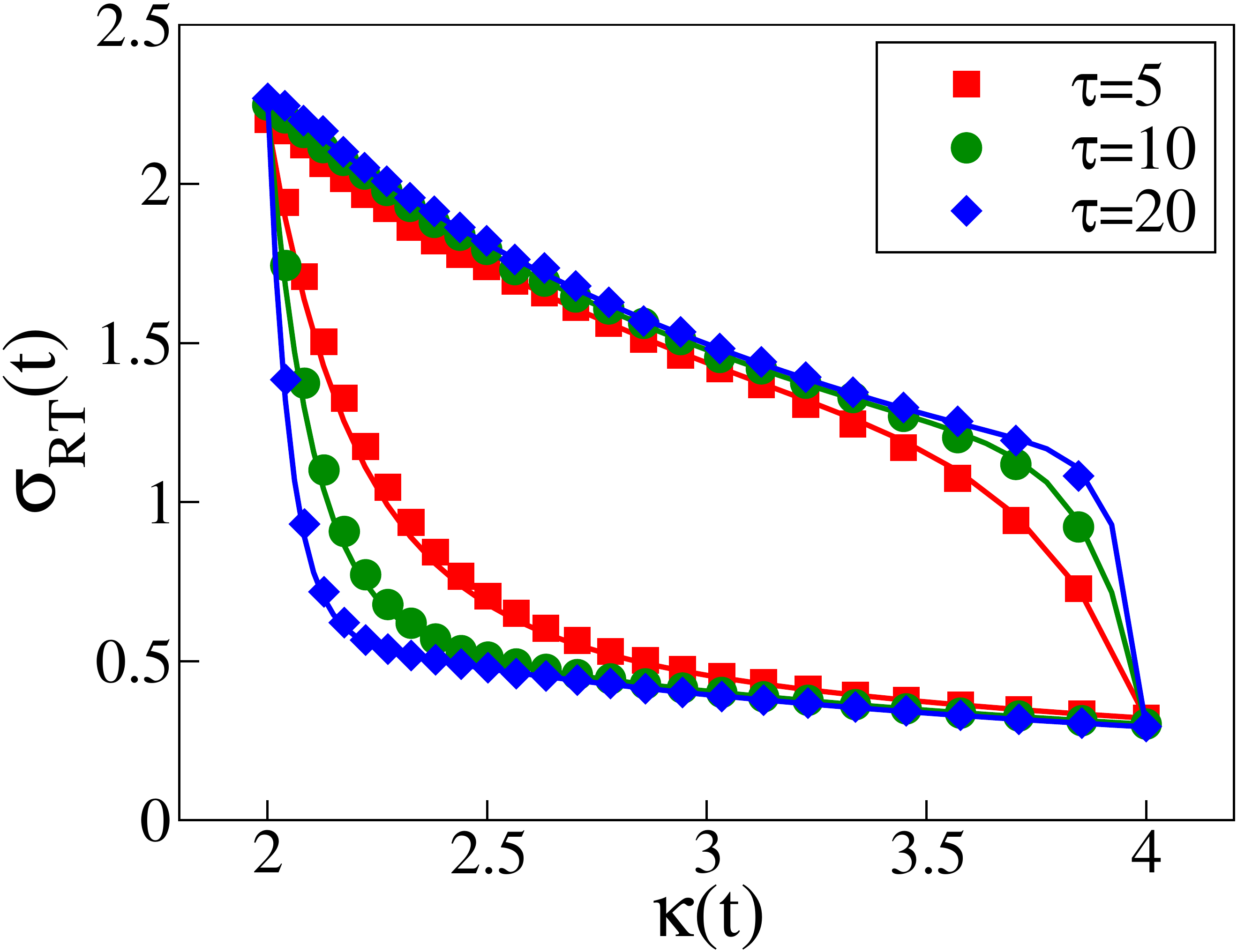}
    \includegraphics[width=8.0cm,height=6.5cm,angle=0]{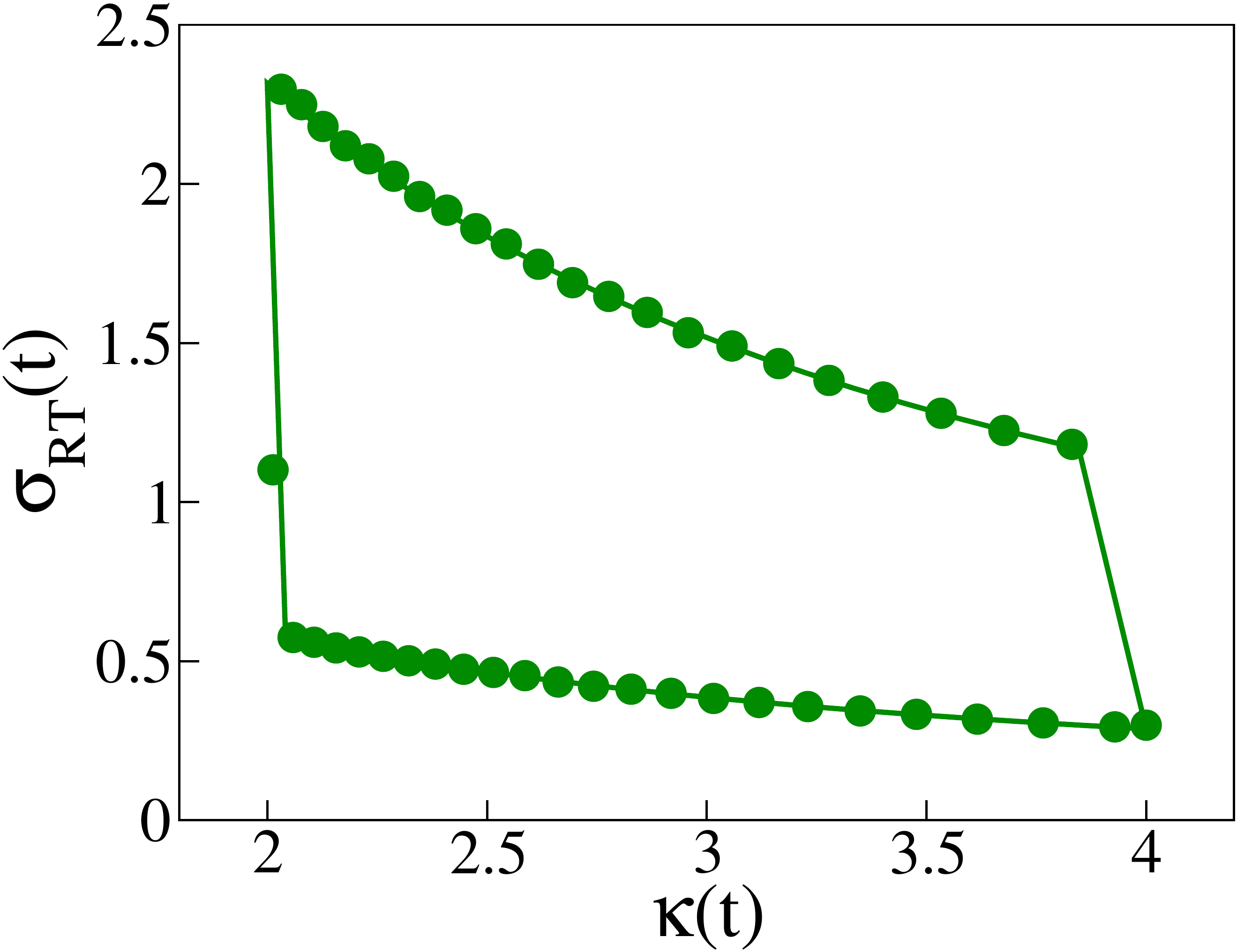}
    \caption{ (Left Panel) simulation and numerical results $\sigma_{RT}(t)$ with $\kappa(t)$ for different $\tau$ for non-quasi static case. (Right panel) result for the quasistatic case. Similar to the passive engine here also we see the four strokes of our engines.  Parameters are $k_{1}=4.0$, $k_{2}=2.0$, $\gamma_{1}=1.0$, $\gamma_{2}=2.0$, $T_{1}=4.0$, $T_{2}=1.0$,  $A_{1}^+=3.0$, $A_{2}^+=2.0$, $\lambda_{1}=5.0$, $\lambda_{2}=5.0$ and $\tau=200$ for the right panel.}
\label{Fig.8}  
\end{figure}
\begin{eqnarray}
	\sigma_{RT}^{(1)}(t)&=&\sigma_{Th}^{(1)}(t)+\frac{\sigma_{RT}(\tau)}{f_{1}^{2}(t)}+\frac{(A_{1}^+)^{2}}{\gamma_{1}^{2}f_{1}^{2}(t)}
	\int_{0}^{t}dt'f_{1}(t')\int_{0}^{t'}dt''e^{-\lambda_{1}( t'-t'')}f_{1}(t'').\nonumber\\
	&&\equiv \sigma_{Th}^{(1)}(t)+\frac{\sigma_{RT}(\tau)}{f_{1}^{2}(t)}+I_{RT}^{(1)}(t)
	\label{Eq.46}
\end{eqnarray}
For the second half of the cycle it becomes
\begin{eqnarray}
\sigma_{RT}^{(2)}(t)&&=\sigma_{Th}^{(2)}(t)+\frac{\sigma_{RT}(\tau/2)}{f_{2}^{2}(t)}+\frac{(A_{2}^+)^{2}}{\gamma_{2}^{2}f_{2}^{2}(t)}\int_{\tau/2}^{t}dt'f_{2}(t')\int_{\tau/2}^{t'}dt''e^{-\lambda_{2}( t'-t'')}f_{2}(t'')\nonumber\\
	&&\equiv \sigma_{Th}^{(2)}(t)+\frac{\sigma_{RT}(\tau/2)}{{f_{2}}^{2}(t)}+I_{RT}^{(2)}(t)
	\label{Eq.47}
\end{eqnarray}
and
\begin{equation}
\sigma_{RT}(\tau)=\frac{I_{RT}^{(1)}(\tau/2)\left(\frac{k_{2}}{k_{1}}\right)^{2a_{2}}+I_{RT}^{(2)}(\tau)}{\left(1-\left(\frac{k_{2}}{k_{1}}\right)^{2(a_{1}+a_{2})}\right)}, ~~  
\sigma_{RT}(\tau/2)=\sigma_{RT}(\tau)\left(\frac{k_{1}}{k_{2}}\right)^{-2a_{1}}+I_{RT}^{(1)}(\tau/2).
\label{sigmaRTs}
\end{equation}
In this case the last integrals in Eqs. (\ref{Eq.46}) and (\ref{Eq.47}) are not exactly solvable for the form of $\kappa(t)$ and hence $f_{1, 2}(t)$ that we have considered. 
We may overcome this by fixing a particular functional form of $f_{1, 2}(t)$ first and then back-calculate the $\kappa(t)$ for the two halves of the cycle. But in this case, from Eq. (\ref{protoform}) it is clear that the protocol $\kappa(t)$ will become a function of the system parameters especially $\gamma_{1, 2}$. This may lead to unphysical situations or may be impossible to achieve in practice. For our case these integrals as well as the ones discussed in the next sections can be solved numerically. 

\begin{figure}[t]
  \hspace{-1cm}
        \includegraphics[width=7.5cm,height=6.5cm,angle=0]{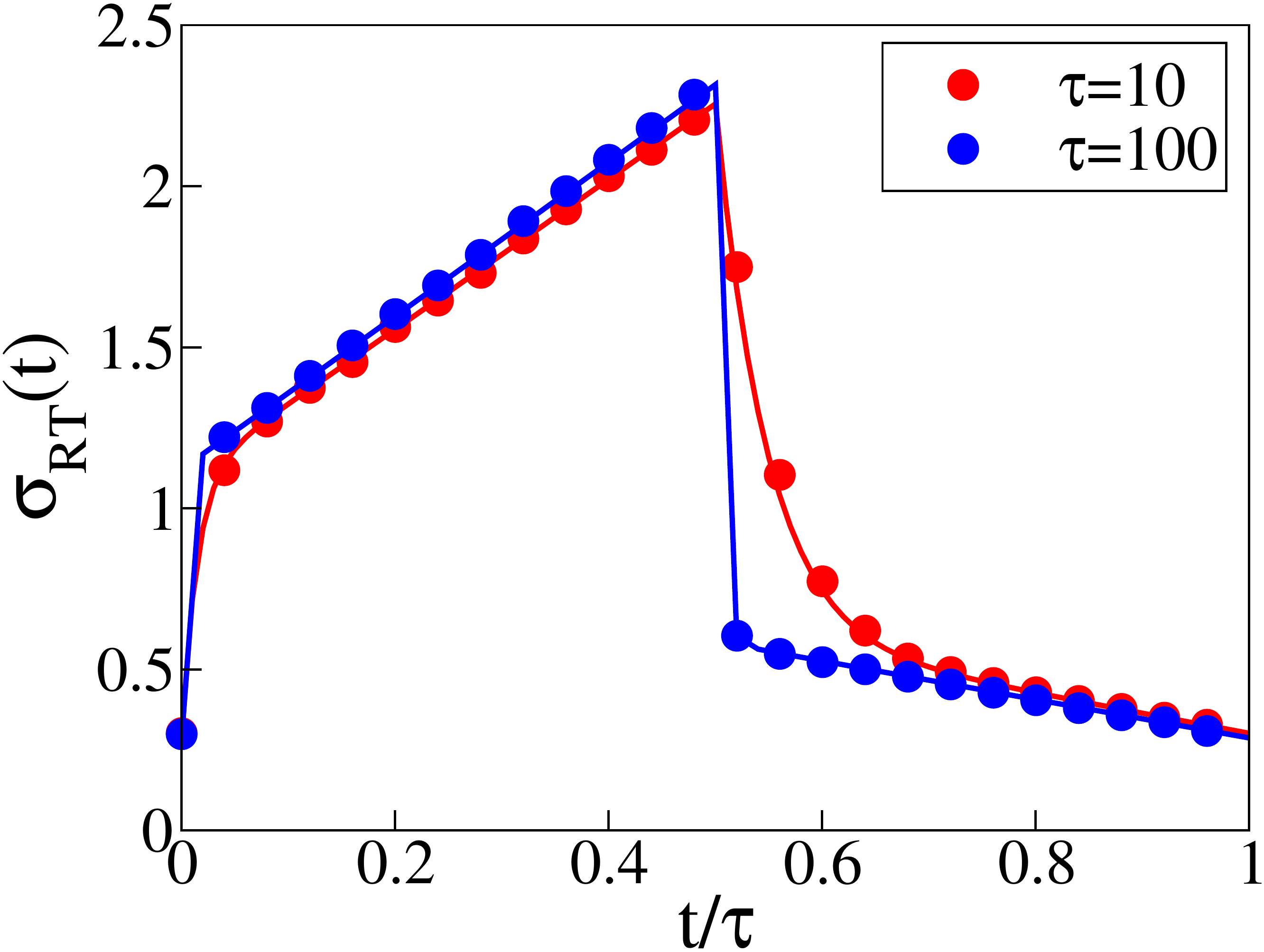}
        \includegraphics[width=7.5cm,height=6.5cm,angle=0]{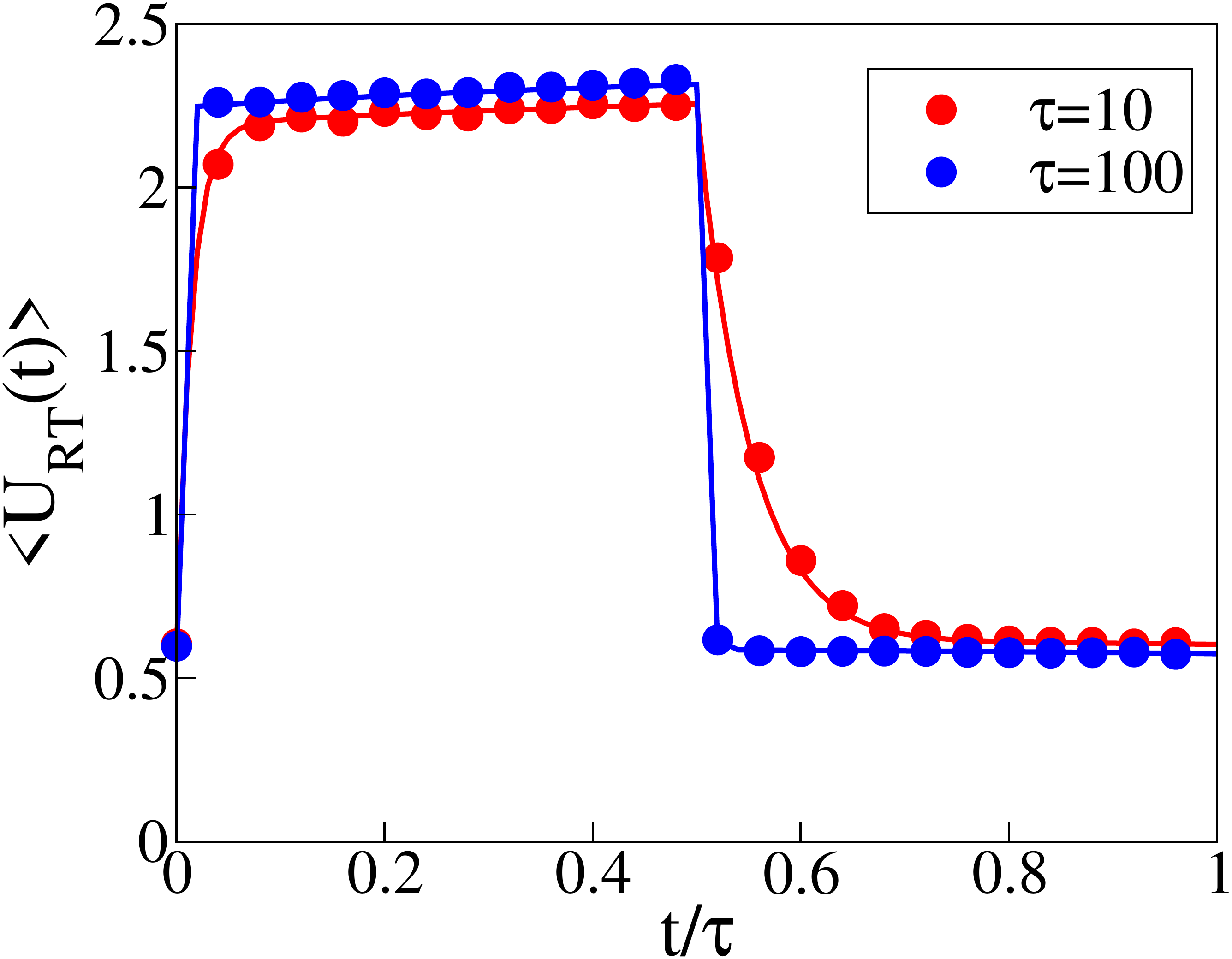}
        \caption{$\sigma_{RT}(t)$ (left panel) and $\langle U(t)\rangle$ (right panel) as a function of $t/\tau$. 
          $k_{1}=4.0$, $ k_{2}=2.0$, $\tau=100$, $\gamma_{1}=1.0$, $\gamma_{2}=2.0$,  $T_{1}=4.0$, $T_{2}=1.0$, $A_{1}^+=3.0$, $A_{2}^+=2.0$, $\lambda_{1}=5.0$, $\lambda_{2}=5.0$}
\label{FIG.11}
\end{figure}

We now present the numerical and simulation results for the RT model. In the Fig. \ref{Fig.8} we have plotted $\sigma_{RT}(t)$ with $\kappa(t)$ for different $\tau$. For large $\tau$ it is similar to the PV diagram of a quasistatic Stirling engine. As earlier, in Fig. {\ref{FIG.11}} we plot the time evolution of $\sigma_{RT}^{(i)}$ and the internal energy for different cycle times.  
We have used the definition of the internal energy same as in the thermal case but one can clearly see that even in the quasistatic limit the equipartition of internal energy is not achieved. Though in this case one may define an effective temperature as   
\begin{equation}
\begin{split}
\langle U_{RT}(t) \rangle =\frac{1}{2}\kappa(t)\langle x_{i}^2\rangle = \frac{k_{B}T^{\text{eff}}_{RT}(t)}{2},
\end{split}
\end{equation}
where $T^{eff}_{RT}(t)$ is the effective temperature experienced by the colloidal particle. This has contributions from the thermal part as well as the part due to the activity namely the athermal part. Thus, $T^{eff}_{RT}(t)=T_{i}+T^{(i)}_{RT}$. And $T^{(i)}_{RT}$ is the athermal part given by \cite{Steffenoni, Wexler20}.
\begin{eqnarray}
T_{RT}^{(1)}(t)&=& \frac{\langle \eta_{RT}^{(1)}(t) x_1(t)\rangle}{k_B} = \frac{(A_{1}^+)^{2}}{2\gamma_{1}f_{1}(t)}\int_{0}^{t}e^{-\lambda_{1}(t-t')}f_{1}(t')dt',\nonumber\\
T_{RT}^{(2)}(t)&=& \frac{\langle \eta_{RT}^{(2)}(t) x_2(t)\rangle}{k_B} =  \frac{(A_{2}^+)^{2}}{2\gamma_{2}f_{2}(t)}\int_{\tau/2}^{t}e^{-\lambda_{1}(t-t')}f_{2}(t')dt'.
\label{TactRT1}
\end{eqnarray}
The internal energy increases with time as the effective temperature increases with the activity. Next we calculate different thermodynamic quantities of interest.

\subsubsection{Calculation of thermodynamic quantities with RT-type activity}
\begin{figure}[!t]
   \hspace{-1cm}
    \includegraphics[width=7.5cm,height=6.0cm,angle=0]{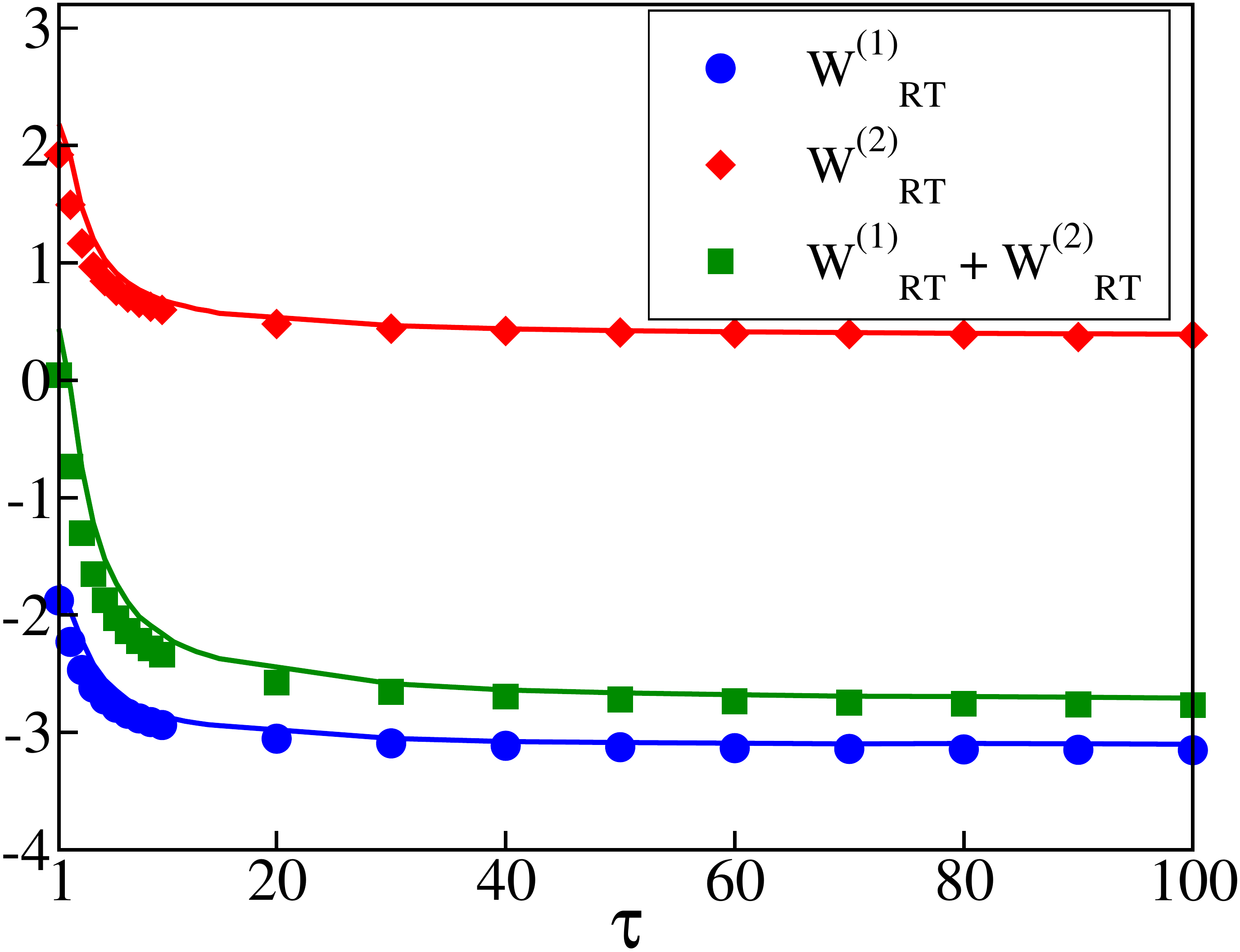}~~~
    \includegraphics[width=7.5cm,height=6.0cm,angle=0]{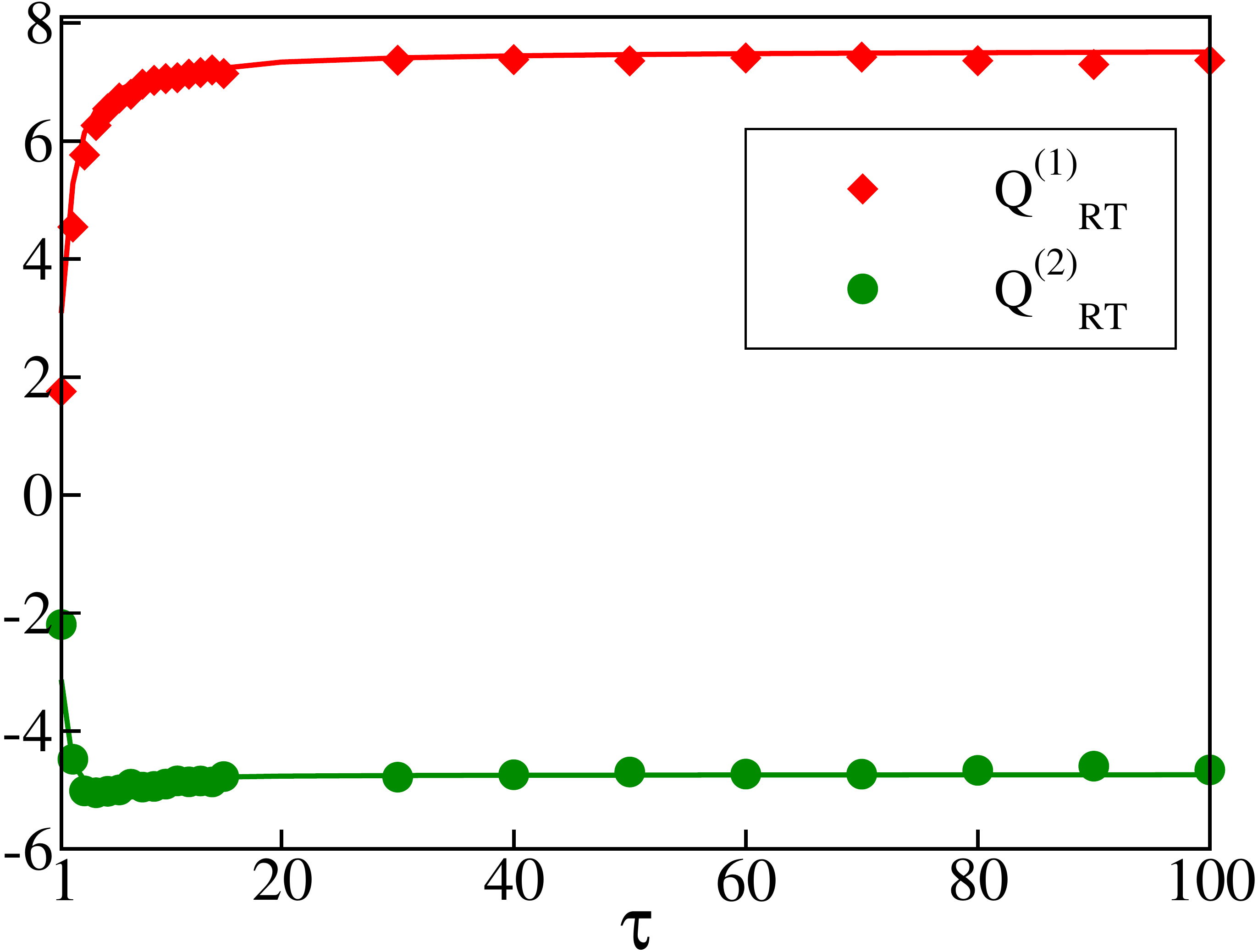}
    \caption{(Left panel) Individual works and total work $ W_{RT}^{(1)} $, $ W_{RT}^{(2)}$ and $W_{RT}^{(1)} + W_{RT}^{(2)}$ with cycle time $\tau$. (Right panel) heats with cycle time $\tau$. Parameters used are $k_{1}=4.0$ , $ k_{2}=2.0$, $\gamma_{1}=1.0$ , $\gamma_{2}=2.0$, $T_{1}=4.0$, $T_{2}=1.0$,  $A_{1}^+=9.0$, $A_{2}^+=1.0$, $\lambda_{1}=5.0 $, $\lambda_{2}=5.0$}
\label{workRTfig}  
\end{figure}
\begin{figure}[!t]
   \hspace{-1cm}
    \includegraphics[width=8.0cm,height=6.0cm,angle=0]{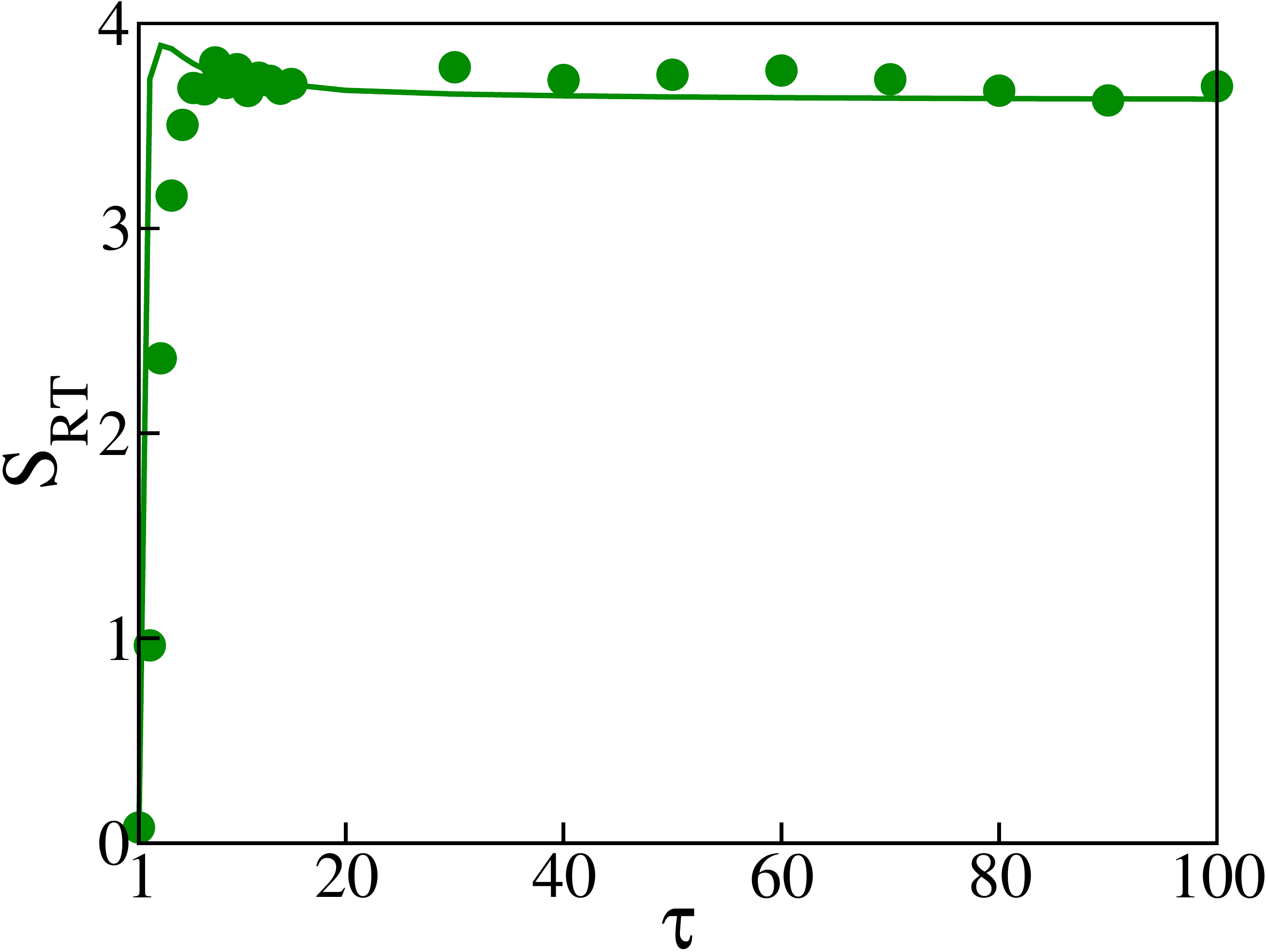}
    \includegraphics[width=8.5cm,height=6.0cm,angle=0]{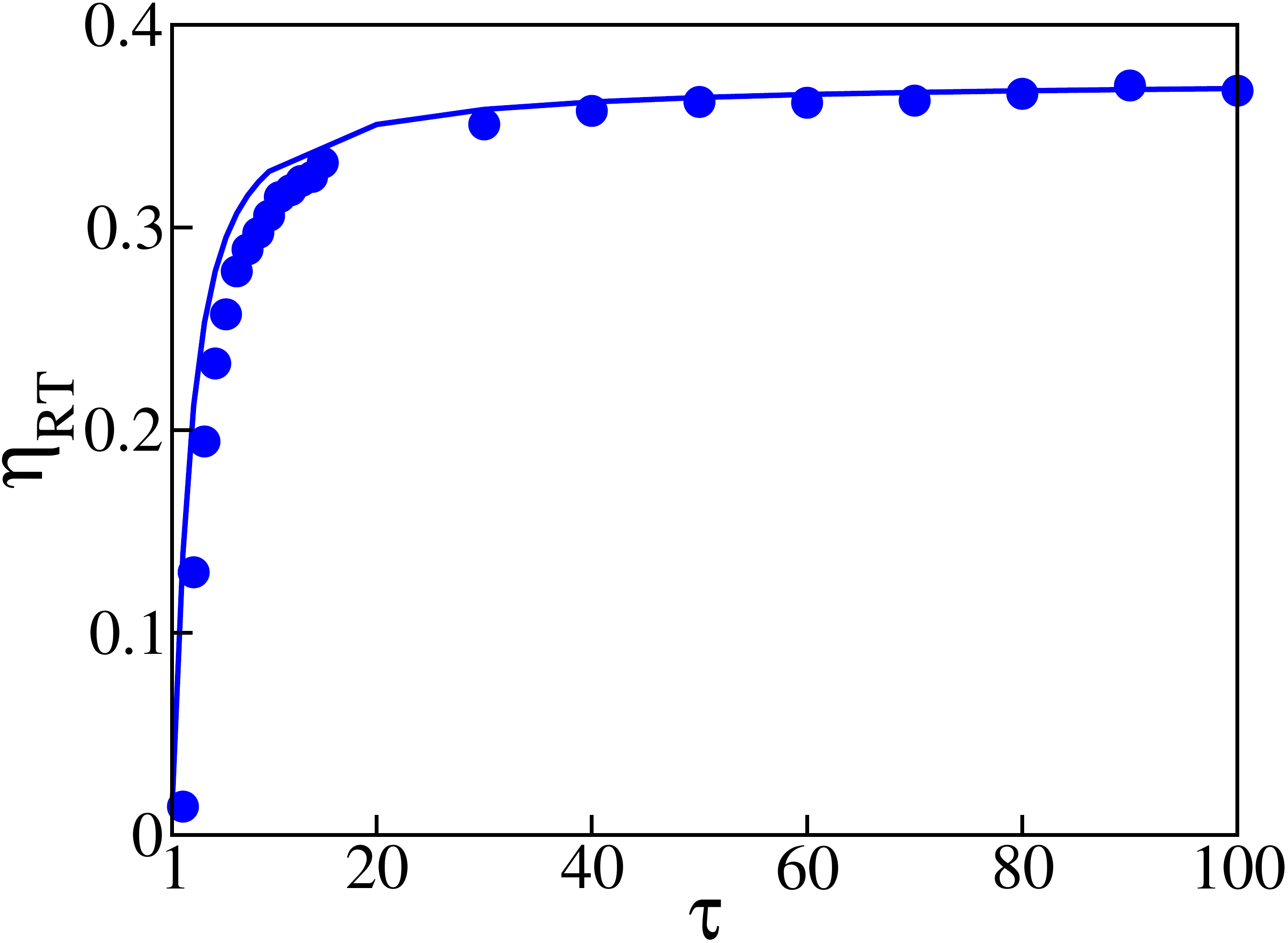}
    \caption{ (Left Panel) Entropy $S_{RT}$ with cycle time $\tau$. Efficiency for the active particle $\eta_{RT}$ with cycle time $\tau$.  Parameters are same as in Fig. \ref{workRTfig}}
\label{entRTfig}  
\end{figure}

We have the expression for $\sigma_{RT}^{(i)}(t)$ (Eqs. (\ref{Eq.46}) (\ref{Eq.47})) for the active case. Hence, using the earlier definition of thermodynamic work we get
\begin{eqnarray}
	W_{RT}^{(1)}(t)&=&W_{Th}^{(1)}(t)+\frac{\sigma_{RT}(\tau)k_{1}}{2(1+2a_{1})}\left[f_{1}(t)^{-\frac{(2a_1+1)}{a_1}}-1\right]-\frac{\gamma_{1}k_{1}(k_{1}-k_{2})}{k_{2}\tau }\int_0^t f_{1}(t')^{-\frac{2}{a_1}}I_{RT}^{(1)}(t')~ dt'\nonumber\\
	\label{Eq.53}
\end{eqnarray}
For the second half of the cycle, the thermodynamic work is
\begin{eqnarray}
W_{RT}^{(2)}(t)&=&W_{Th}^{(2)}(t)+\frac{\sigma_{RT}(\frac{\tau}{2})k_{2}}{2(1-2a_{2})}\left[f_2(t)^{-\frac{(2a_2-1)}{a_2}}-1\right]+\frac{\gamma_{2}k_{2}(k_1-k_2)}{k_{1}\tau}\int_{\tau/2}^{t} f_{2}(t')^{\frac{2}{a_{2}}}I_{RT}^{(2)}(t')~ dt',\nonumber\\
	\label{Eq.55}
\end{eqnarray}
The expressions for the heat in first and second half are
\begin{eqnarray}
&&Q_{RT}^{(1)}=\frac{1}{2}\left[\kappa_{1}(\frac{\tau}{2})\sigma^{(1)}_{RT}(\frac{\tau}{2})-\kappa_{1}(0)\sigma^{(1)}_{RT}(0)\right]-W^{(1)}_{RT}(\frac{\tau}{2})
\label{Eq.32rt}
\end{eqnarray}
and during the contraction step we get
\begin{eqnarray}
Q_{RT}^{(2)}&=&\frac{1}{2}\left[\kappa_{2}(\tau)\sigma^{(1)}_{RT}(\tau)-\kappa_{2}(\frac{\tau}{2})\sigma^{(2)}_{RT}(\frac{\tau}{2})\right]-W^{(2)}_{RT}(\tau)
\label{Eq.33rt}
\end{eqnarray}
\begin{equation}
S_{RT}^{(i)}(t)=-\int^{t}\frac{1}{T_{i}+T_{RT}^{(i)}(t')}\left(\frac{\dot f_i(t')}{f_i(t')}\right)\left(-\gamma_{i} \left(\frac{\dot f_i(t')}{f_i(t')}\right)\sigma^{(i)}_{RT}(t')+k_{B} T_{i}+k_{B}T^{(i)}_{RT}(t')\right)dt'.
\label{entropyRT}  
\end{equation}
Using $Q_{RT}^{(1)}$ and $W_{RT}^{(i)}$ we calculate the efficiency, as it is defined earlier. Work, heat, entropy and efficiency are plotted in Fig. \ref{workRTfig} and \ref{entRTfig} from analytics as well as from simulation and they match well.

\subsubsection{Thermodynamics of active micro heat engine with Gaussian coloured noise}
So far, we have discussed the non-Gaussian active noise (RT) and calculated all thermodynamic quantities. Here, we calculate thermodynamics of active single-particle micro heat engine with another simplified model of active fluctuations namely Active Ornstein-Uhlenbeck particles (AOUP). Unlike RT, this noise follows Gaussian statistics and is exponentially correlated. Similar to RT noise this also breaks the FDR. The AOUP process is modeled by writing another Lagevin equation for the noise $\eta_{CN}(t)$ itself along with the Langevin equation (Eq. (\ref{eom})) as 
\begin{equation}
\tau_{a}\dot{\eta}^{(i)}_{CN}(t)~=~-\eta^{(i)}_{CN}(t)~+~\zeta_{i}(t),
\label{Eq.100}
\end{equation}
where $\zeta_{i}$ is the delta correlated Gaussian white noise with zero mean and variance $D_i$. This gives us the required exponentially correlated colored noise with the correlation time $\tau_{a}$ namely
\begin{eqnarray}
\langle\eta^{(i)}_{CN}(t)\eta^{(i)}_{CN}(t')\rangle&=&\frac{D_{i}}{\tau_{a}}e^{-\frac{\vert t-t'\vert}{\tau_{a}}}.
\label{CN}
\end{eqnarray}
Here CN stands for the ``Colored Noise''. The constant $D_i$ is proportional to the strength of the correlation and $\tau_a$ associated with the persistence of the active colloidal particle which is introduced in its dynamics by the incessant interaction between the colloidal degree of freedom and active degrees of freedom of the reservoir. It corresponds to the correlation time scale between two consecutive active kicks on the colloidal particle by the active particle. If we consider $\frac{(A_i^+)^2}{2}=\lambda_i D_i=D_i/\tau_a$ and $\lambda_i=\tau_a^{-1}$, then the AOUP converges to the RT model. For AOUP, if  $\frac{D_i}{\tau_a}=2\gamma_ik_BT_i$ , in the $\tau_a\rightarrow 0$ limit we have $\langle\eta_i(t)\eta_j(t')\rangle \rightarrow \langle\xi_i(t)\xi_j(t')\rangle$. This implies that the thermal noise can be recovered from AOUP noise in this limit. The finite value of $\tau_{a}$ implies finite activity and zero $\tau_{a}$, the activity vanishes. 
\subsubsection{Time evolution of the position variance of the particle under AOUP}
The variance of the position of the colloid 
For this we use Eq. (\ref{formal1}, \ref{sigma}) with both the thermal and active noise. Other details of the system are identical to the ones in the passive micro heat engine. The variance is,

\begin{figure}[t]
	\hspace{-1cm}
	\includegraphics[width=8.0cm,height=6.5cm,angle=0]{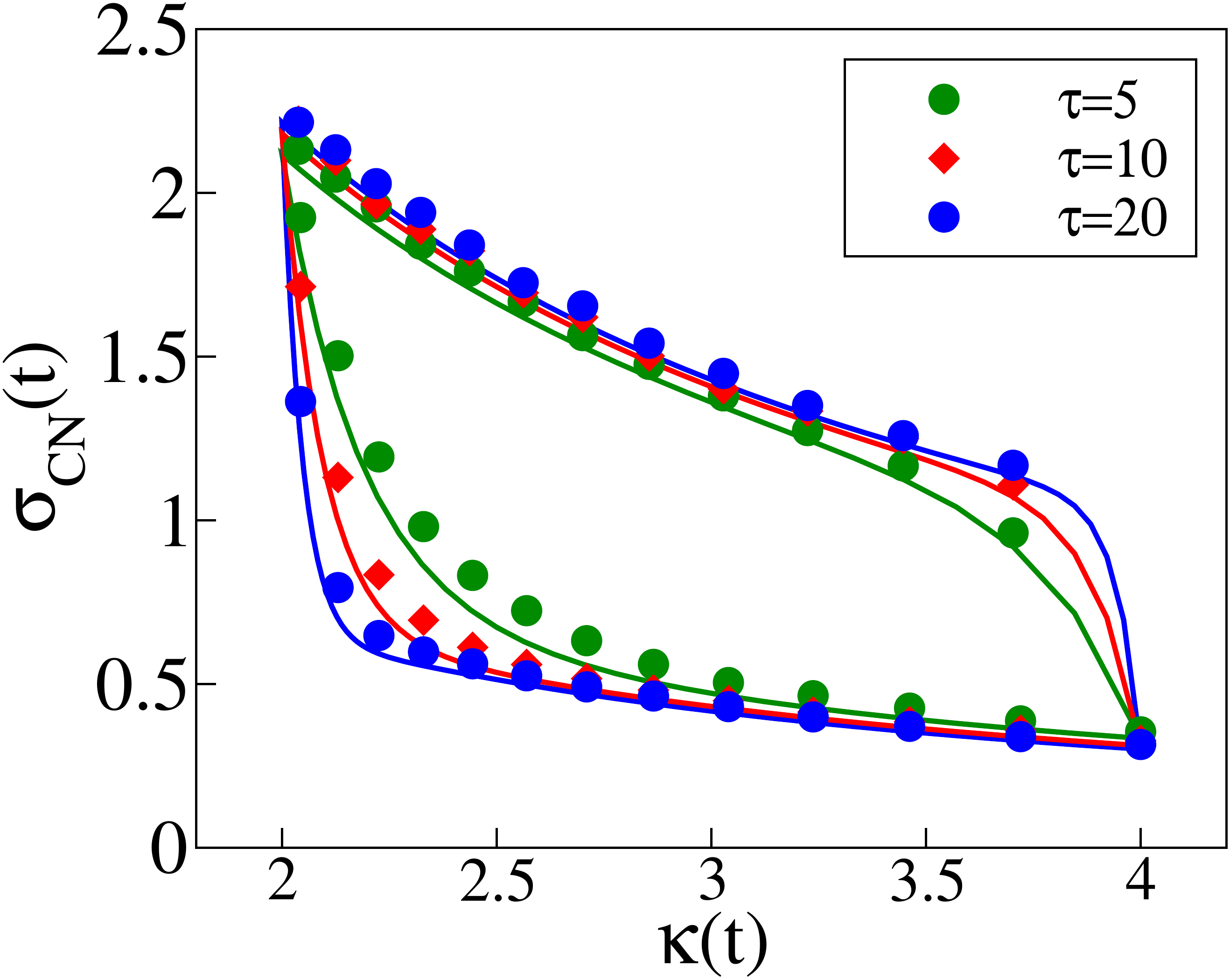}
	\includegraphics[width=8.0cm,height=6.5cm,angle=0]{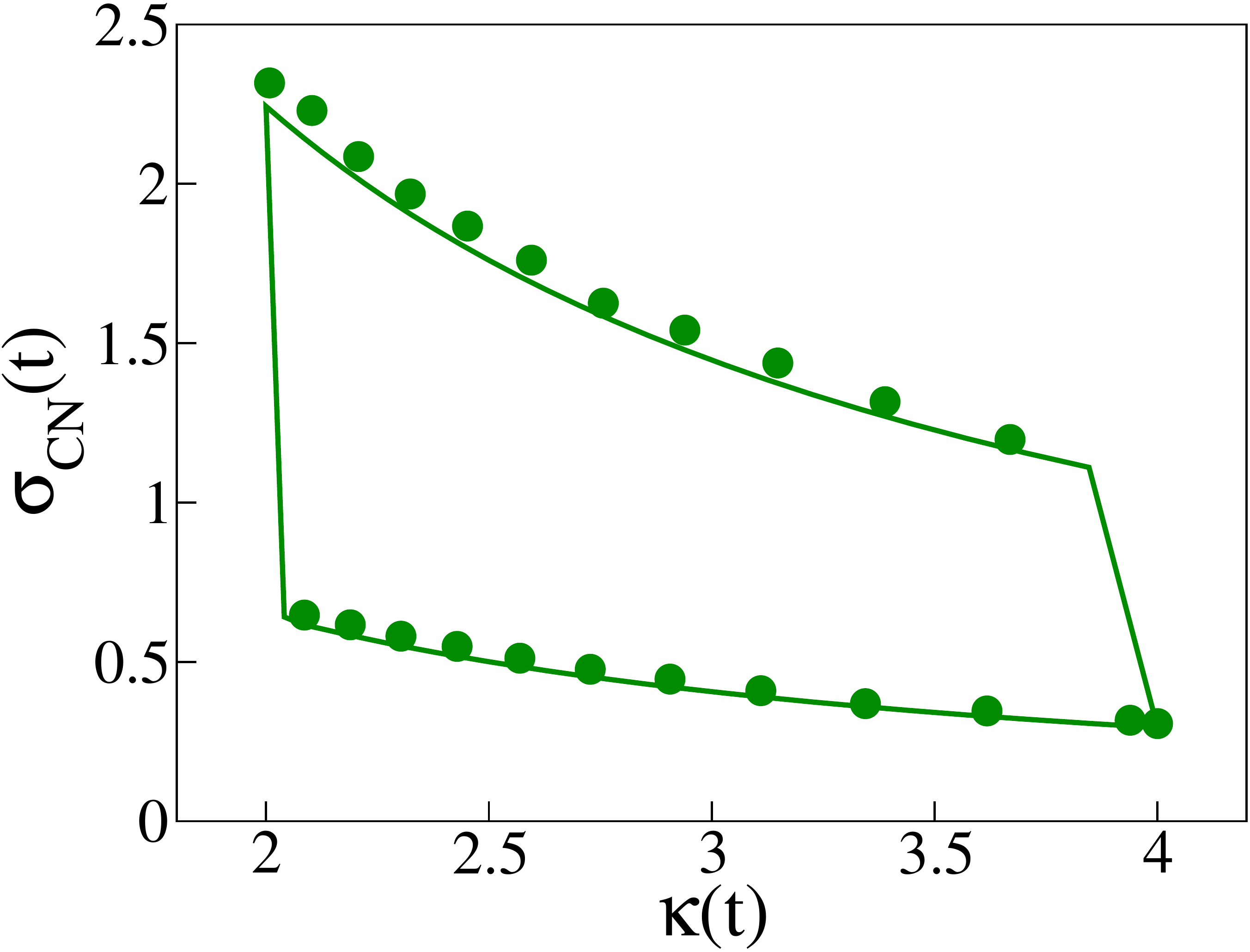}
	\caption{ (Left Panel) simulation and numerical results $\sigma_{CN}(t)$ with $\kappa(t)$ for different $\tau$ for non-quasi static case. (Right panel) result for the quasistatic case. Similar to the passive engine here also we see the four strokes of our engines.  Parameters are $k_{1}=4.0$, $k_{2}=2.0$, $\gamma_{1}=1.0$, $\gamma_{2}=2.0$, $T_{1}=4.0$, $T_{2}=1.0$ $D_{1}=3.0$, $D_{2}=2.0$, $\tau_{a}=2.5$ and $\tau=200$ for the right panel.}
	\label{Fig.12}  
\end{figure}
\begin{eqnarray}
	\sigma_{CN}^{(1)}(t)&=&\sigma_{Th}^{(1)}(t)+\frac{\sigma_{CN}(\tau)}{f_{1}^{2}(t)}+\frac{2D_{1}}{\tau_{a}\gamma_1^2f_{1}^{2}(t)}\int_{0}^{t}dt'f_{1}(t')\int_{0}^{t'}dt''e^{-\frac{(t'-t'')}{\tau_{a}}}f_{1}(t'')\nonumber\\
	&\equiv& \sigma_{Th}^{(1)}(t)+\frac{\sigma_{CN}(\tau)}{f_{1}^{2}(t)}+I_{CN}^{(1)}(t)
	\label{Eq.68}
\end{eqnarray}
For the second half of the cycle it becomes,
\begin{eqnarray}
	\sigma_{CN}^{(2)}(t)&=&\sigma_{Th}^{(2)}(t)+\frac{\sigma_{CN}(\tau/2)}{f_{2}^{2}(t)}+\frac{2D_{2}}{\tau_{a}\gamma_{2}^{2}f_{2}^{2}(t)}\int_{\tau/2}^{t}dt'f_{2}(t')\int_{\tau/2}^{t'}dt''e^{-\frac{(t'-t'')}{\tau_{a}}}f_{2}(t'')\nonumber\\
	&\equiv & \sigma_{Th}^{(2)}(t)+\frac{\sigma_{CN}(\tau/2)}{f_{2}^{2}(t)}+ I_{CN}^{(2)}(t')
	\label{Eq.69}
\end{eqnarray}
and
\begin{equation}
\sigma_{CN}(\tau)=\frac{I_{CN}^{(1)}(\tau/2)\left(\frac{k_{2}}{k_{1}}\right)^{2a_{2}}+I_{CN}^{(2)}(\tau)}{\left(1-\left(\frac{k_{2}}{k_{1}}\right)^{2(a_{1}+a_{2})}\right)}, ~~  
\sigma_{CN}(\tau/2)=\sigma_{CN}(\tau)\left(\frac{k_{1}}{k_{2}}\right)^{-2a_{1}}+I_{CN}^{(1)}(\tau/2).
\label{sigmaCNs}
\end{equation}
In Fig. \ref{Fig.12}, we have plotted $ \sigma_{CN}(t)$ with $\kappa(t)$ which is again similar to the PV diagram of a Stirling engine. Similar to thermal case we now plot the simulation and analytical results for the time evolution of $\sigma_{CN}^{(i)}$ and the internal energy for different cycle times in Fig. \ref{Fig.13}.
\begin{figure}[t]
	\hspace{-1cm}
	\includegraphics[width=7.5cm,height=6.5cm,angle=0]{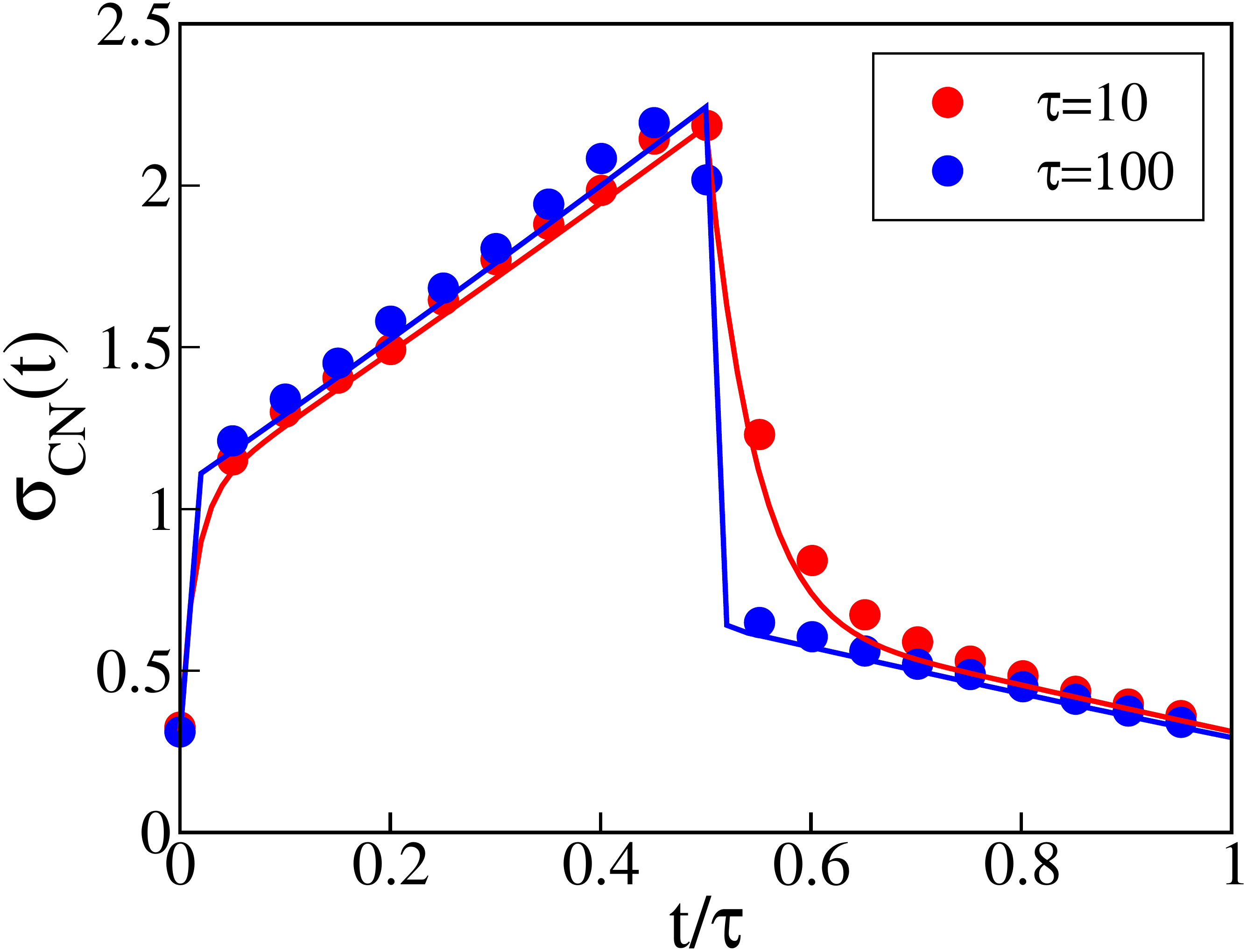}
	\includegraphics[width=7.5cm,height=6.5cm,angle=0]{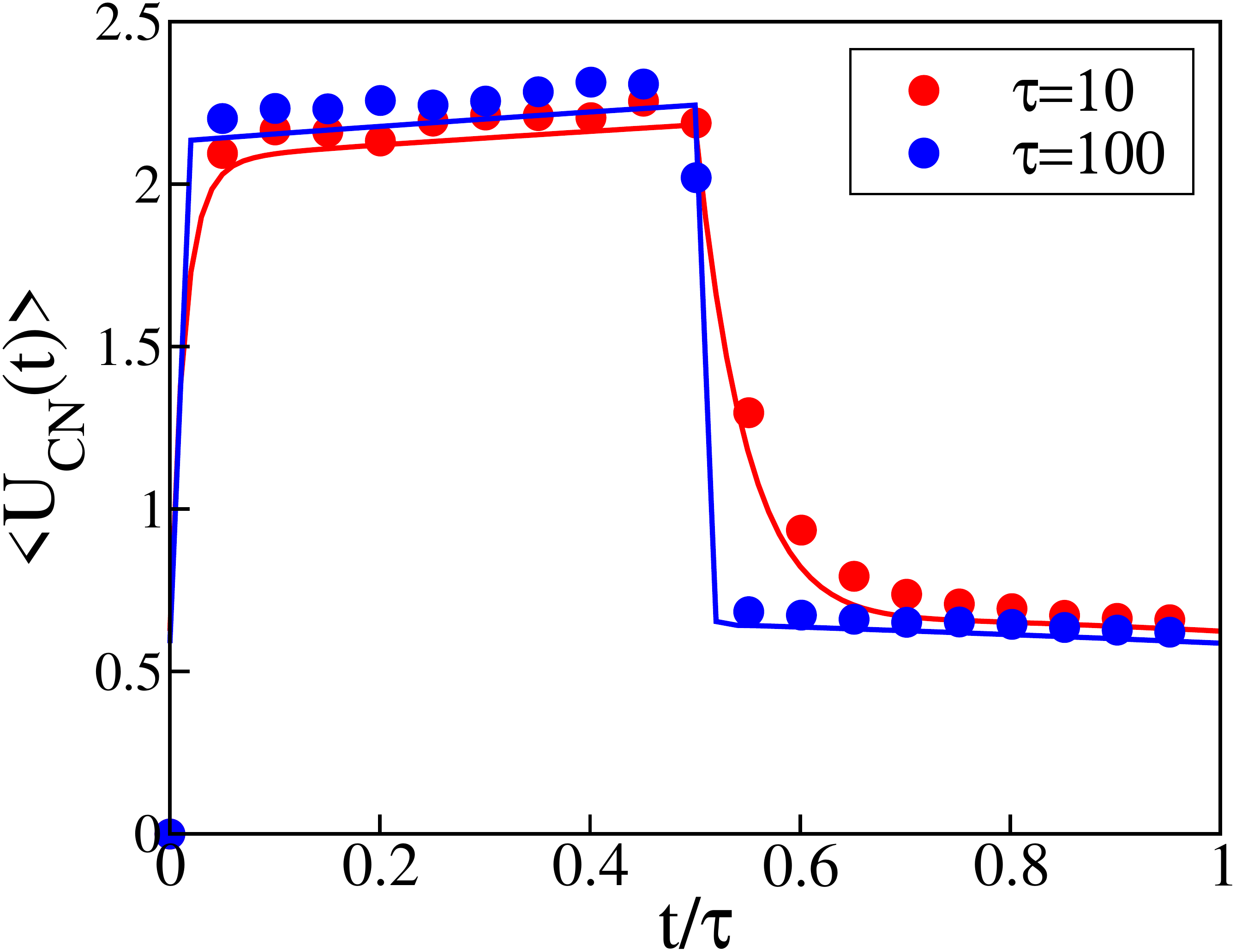}
	\caption{$\sigma_{CN}(t)$ (left panel) and $\langle U_{CN}(t)\rangle$ (right panel) as a function of $t/\tau$. Parameters are 
		$k_{1}=4.0$, $ k_{2}=2.0$, $\tau=100$, $\gamma_{1}=1.0$, $\gamma_{2}=2.0$, $T_{1}=4.0$, $T_{2}=1.0$, $D_{1}=3.0$, $D_{2}=2.0$, $\tau_{a}=2.5$.}
	\label{Fig.13}
\end{figure}
We have used the definition of the internal energy as in the thermal case but we see that in the quasistatic limit the equipartition of internal energy does not seem to be achieved. Although the matching is much better than the RT case. We then use time dependent effective temperature $T^{\text{eff}}_{CN}(t)$ as done earlier
\begin{equation}
\begin{split}
\langle U_{CN}(t) \rangle =\frac{1}{2}\kappa(t)\langle x_{i}^2\rangle = \frac{k_{B}T^{\text{eff}}_{CN}(t)}{2},
\end{split}
\end{equation}
It has contributions from the thermal part as well as from the the active degrees of freedom i.e. the athermal part. Thus the internal energy increases as the effective temperature increases with the activity of the bath. 

\subsubsection{Thermodynamic quantities with coloured noise}
We have the expression for $\sigma_{CN}^{(i)}(t)$ for the active case. Hence, using the definition of thermodynamic work we get,
\begin{eqnarray}
	&&W_{CN}^{(1)}(t)=W_{Th}^{(1)}(t)+\frac{\sigma_{CN}(\tau)k_{1}}{2(1+2a_{1})}\left[f_{1}(t)^{-\frac{(1+2a_1)}{a_1}}-1\right]-\frac{\gamma_{1}k_{1}(k_{1}-k_{2})}{\tau k_{2}}\int_0^t f_1(t')^{-\frac{2}{a_1}} I_{CN}^{(1)}(t')dt'\nonumber\\
	\label{Eq.72}
\end{eqnarray}
Similarly for the second half of the cycle, the thermodynamic work is
\begin{eqnarray}
	W_{CN}^{(2)}(t)&=&W_{Th}^{(2)}(t)+\frac{\sigma_{CN}(\frac{\tau}{2})k_{2}}{2(1-2a_{2})}\left[f_{1}(t)^{-\frac{(2a_2-1)}{a_2}}-1\right]+\frac{\gamma_{2}k_{2}(k_{1}-k_{2})}{k_{1}\tau}\int_{\tau/2}^{t}f_2(t')^{\frac{2}{a_2}}I_{CN}^{(2)}(t'),\nonumber\\
	\label{Eq.73}
\end{eqnarray}
The expression for the heat in first half is
\begin{eqnarray}
&&Q_{CN}^{(1)}=\frac{1}{2}\left[\kappa_{1}(\frac{\tau}{2})\sigma^{(1)}_{CN}(\frac{\tau}{2})-\kappa_{1}(0)\sigma^{(1)}_{CN}(0)\right]-W^{(1)}_{CN}(\frac{\tau}{2})
\label{Eq.32cn}
\end{eqnarray}
and for the second half it becomes
\begin{eqnarray}
Q_{CN}^{(2)}&=&\frac{1}{2}\left[\kappa_{2}(\tau)\sigma^{(1)}_{CN}(\tau)-\kappa_{2}(\frac{\tau}{2})\sigma^{(2)}_{CN}(\frac{\tau}{2})\right]-W^{(2)}_{CN}(\tau)
\label{Eq.33cn}
\end{eqnarray}
The entropy in different halves of the cycle is 
\begin{equation}
S_{CN}^{(i)}(t)=-\int^{t}\frac{1}{T_{i}+T_{CN}^{(i)}(t')}\left(\frac{\dot f_i(t')}{f_i(t')}\right)\left(-\gamma_{i} \left(\frac{\dot f_i(t')}{f_i(t')}\right)\sigma^{(i)}_{CN}(t')+k_{B} T_{i}+k_{B}T^{(i)}_{CN}(t')\right)dt'.
\label{entropyCN}  
\end{equation}
Here, the expression for $T_{CN}^{(i)}$ are given by
\begin{eqnarray}
T_{CN}^{(1)}(t)=\frac{D_{1}}{\tau_{a}\gamma_{1}f_{1}(t)}\int_{0}^{t}e^{-\frac{(t-t')}{\tau_{a}}}f_{1}(t')dt',\nonumber\\
T_{CN}^{(2)}(t)=\frac{D_{2}}{\tau_{a}\gamma_{2}f_{2}(t)}\int_{\tau/2}^{t}e^{-\frac{(t-t')}{\tau_{a}}}f_{2}(t')dt'.
\label{TactCN1}
\end{eqnarray}
Using $Q_{CN}^{(1)}$ and $W_{CN}^{(i)}$ we calculate the efficiency, as defined earlier. Work, heat, entropy and efficiency are plotted in Fig. {\ref{workCNfig} and \ref{entCNfig} from analytics as well as from simulation and they match well.
\begin{figure}[!t]
	\hspace{-1cm}
	\includegraphics[width=7.5cm,height=6.0cm,angle=0]{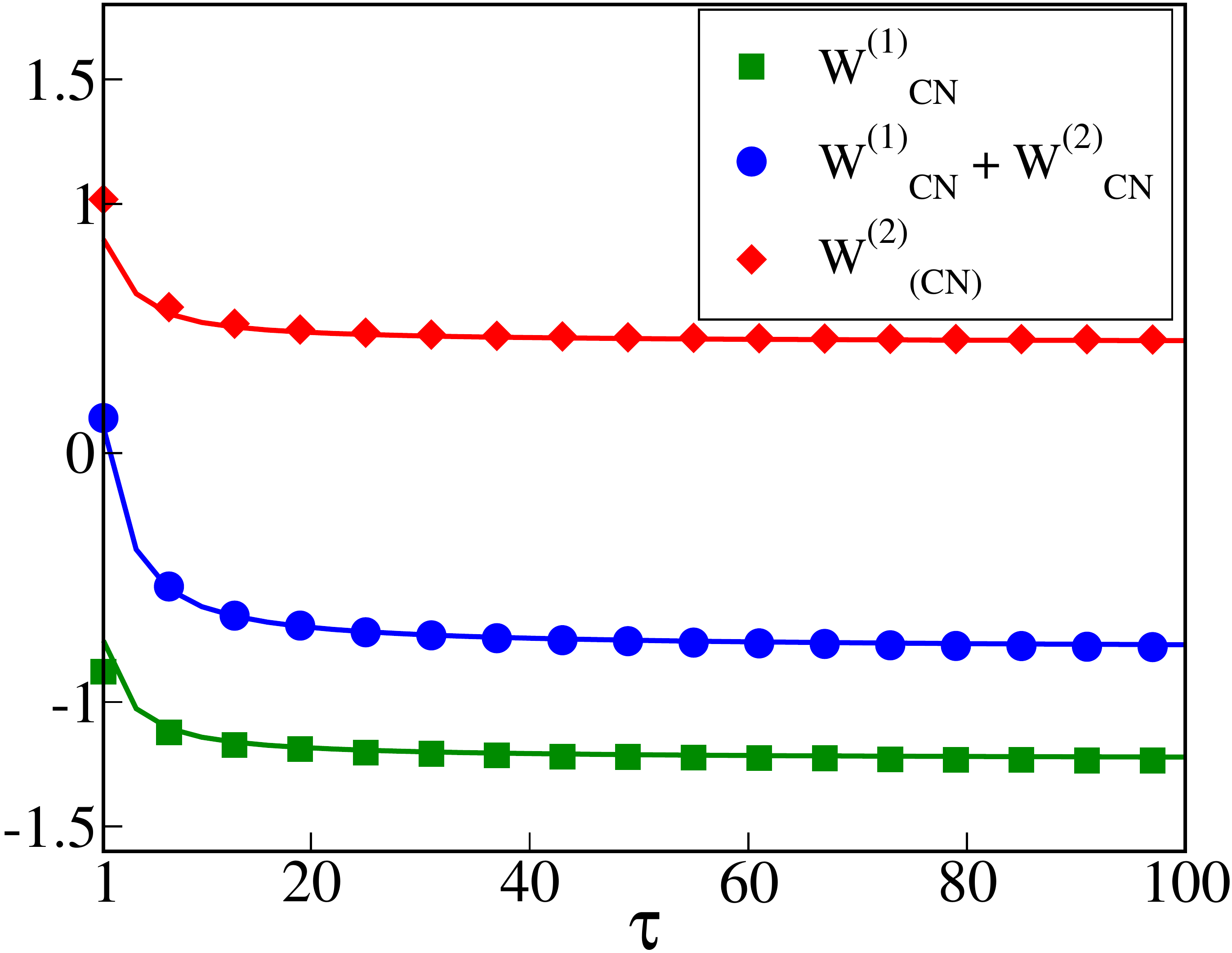}~~~
	\includegraphics[width=7.5cm,height=6.0cm,angle=0]{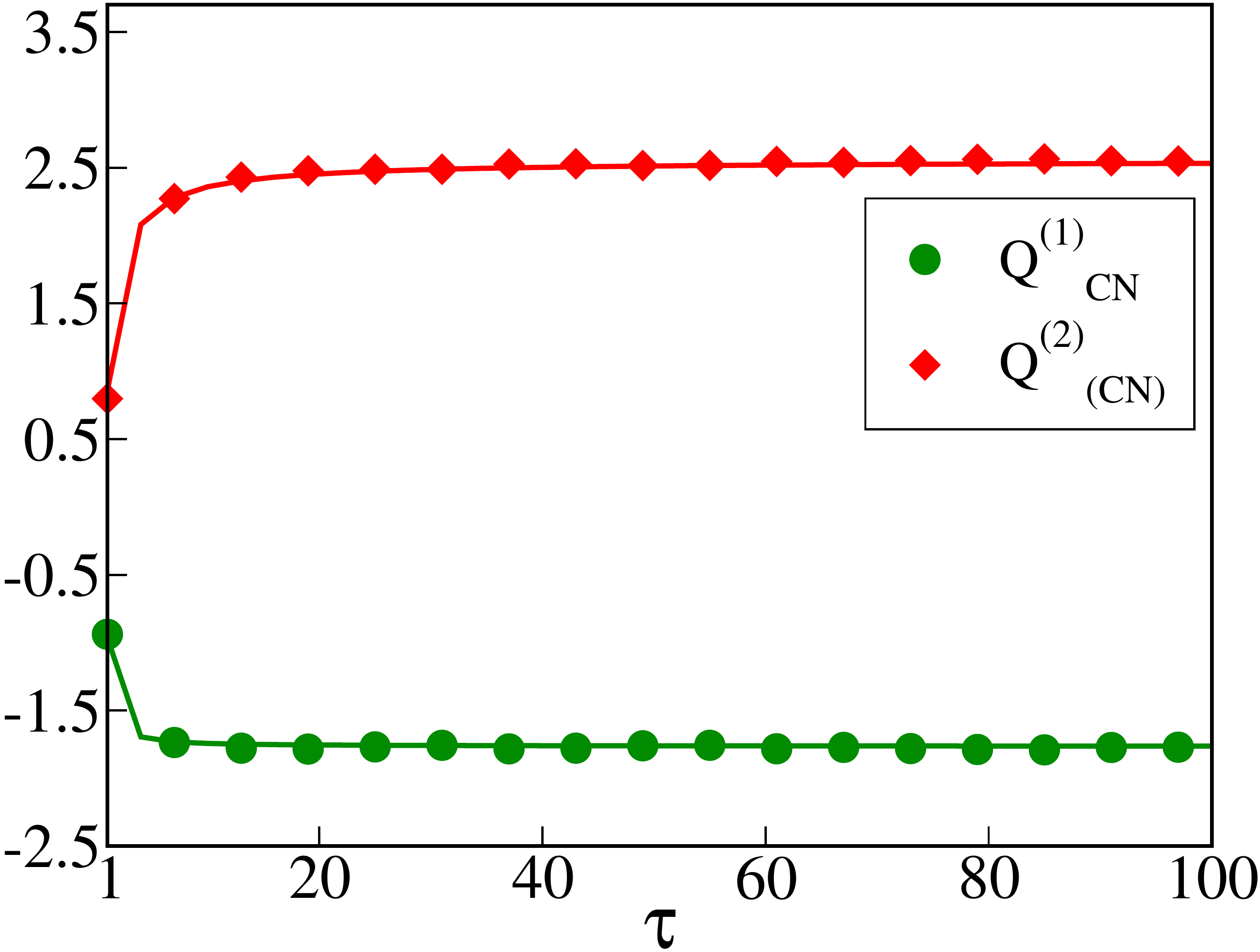}
	\caption{(Left panel) Individual works and total work $ W_{CN}^{(1)} $, $ W_{CN}^{(2)}$ and $W_{CN}^{(1)} + W_{RT}^{(2)}$ with cycle time $\tau$. (Right panel) heats with cycle time $\tau$. Parameters used are $k_{1}=4.0$, $ k_{2}=2.0$, $\gamma_{1}=1.0$, $\gamma_{2}=2.0$, $T_{1}=1.0$, $T_{2}=0.5$,  $D_{1}=4.0$, $D_{2}=2.0$, $\tau_{a}=0.2$}
	\label{workCNfig}  
\end{figure}
\begin{figure}[!t]
	\hspace{-1cm}
	\includegraphics[width=8.0cm,height=6.0cm,angle=0]{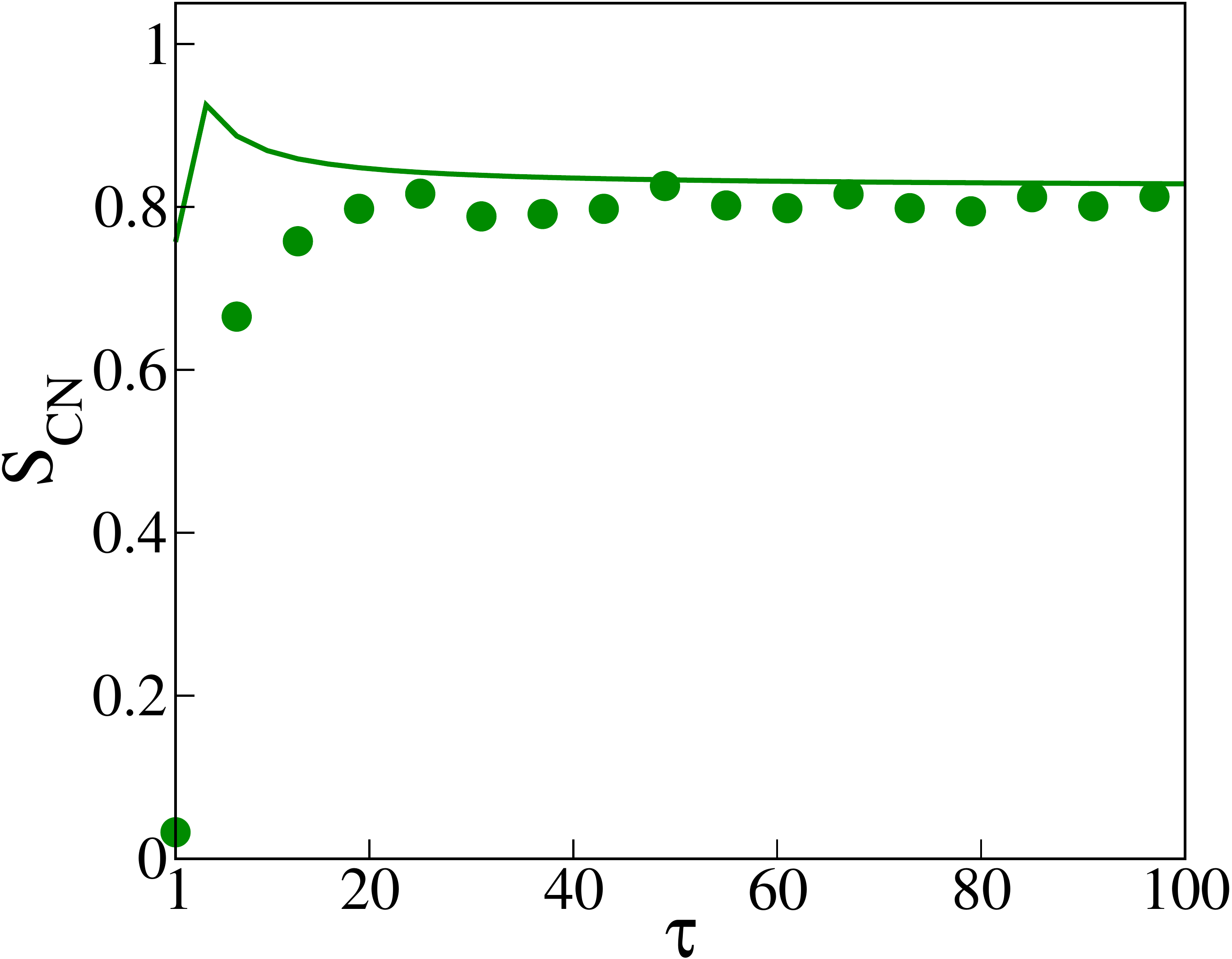}
	\includegraphics[width=8.5cm,height=6.0cm,angle=0]{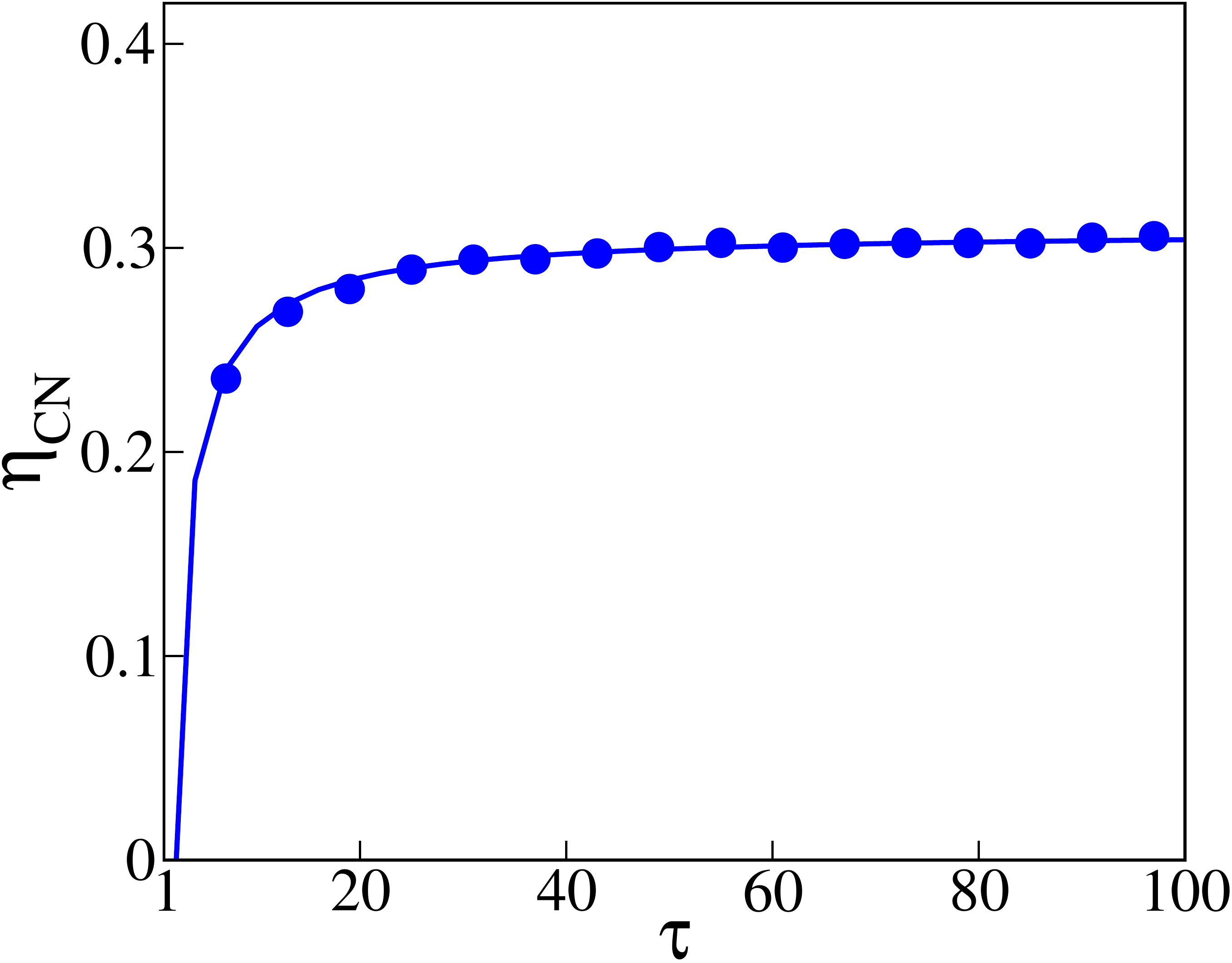}
	\caption{ (Left Panel) Entropy $S_{CN}$ with cycle time $\tau$. Efficiency for the active particle $\eta_{CN}$ with cycle time $\tau$.  Parameters are same as in Fig. \ref{workCNfig}}
	\label{entCNfig}  
\end{figure}
\section{Comparative study of the micro heat engines}
We now compare the performance of the thermal engine with the active engines. For this we introduce a time dependent parameter $\beta=\frac{\lambda_{i}}{\kappa_{i}(t)}$ to compare the characteristics of passive and active models of micro heat engines \cite{urna20}. It essentially signifies the competition between two opposite effects: propulsion of the particle due to the correlated/persistent kicks of surrounding active particles, for which the particle is moving away from its initial position (say the mean position of the harmonic trap) and the harmonic trapping itself for which the particle is attracted towards the mean position of the trap. If the propulsion due to activity dominates, then $\beta < 1$. Whereas, if the harmonic trapping dominates then $\beta>1$.  Thus $\beta$ turns out to be an important parameter to distinguish the characteristics of passive and active micro-heat engines. In recent studies it was also observed that the active engines can give rise to higher efficiencies than the passive engines \cite{Krishnamurty16, Arnab19}. The reason was associated to the non-Gaussian distributions with long tails of the probability distributions of positions of the trapped particle. To compare the steady state distributions of particle's position in various cases, we collect the data at a particular time at $t=\frac{\tau}{4}$ during the protocol. The parameter $\beta$ has the corresponding value at that time. The measure of non-Gaussianity can be calculated by excess kurtosis defined as $\kappa_{exss}= \frac{\langle (x-\langle x \rangle)^4\rangle}{(\langle (x-\langle x \rangle)^2\rangle)^2}-3.0$. For Gaussian distributions $\kappa_{exss}=0$ and it can be greater than or less than zero depending on how slow or fast the tails of distributions decay as compared to the Gaussian distribution. 

\begin{figure}[!t]
    \includegraphics[width=8cm,height=6.0cm,angle=0]{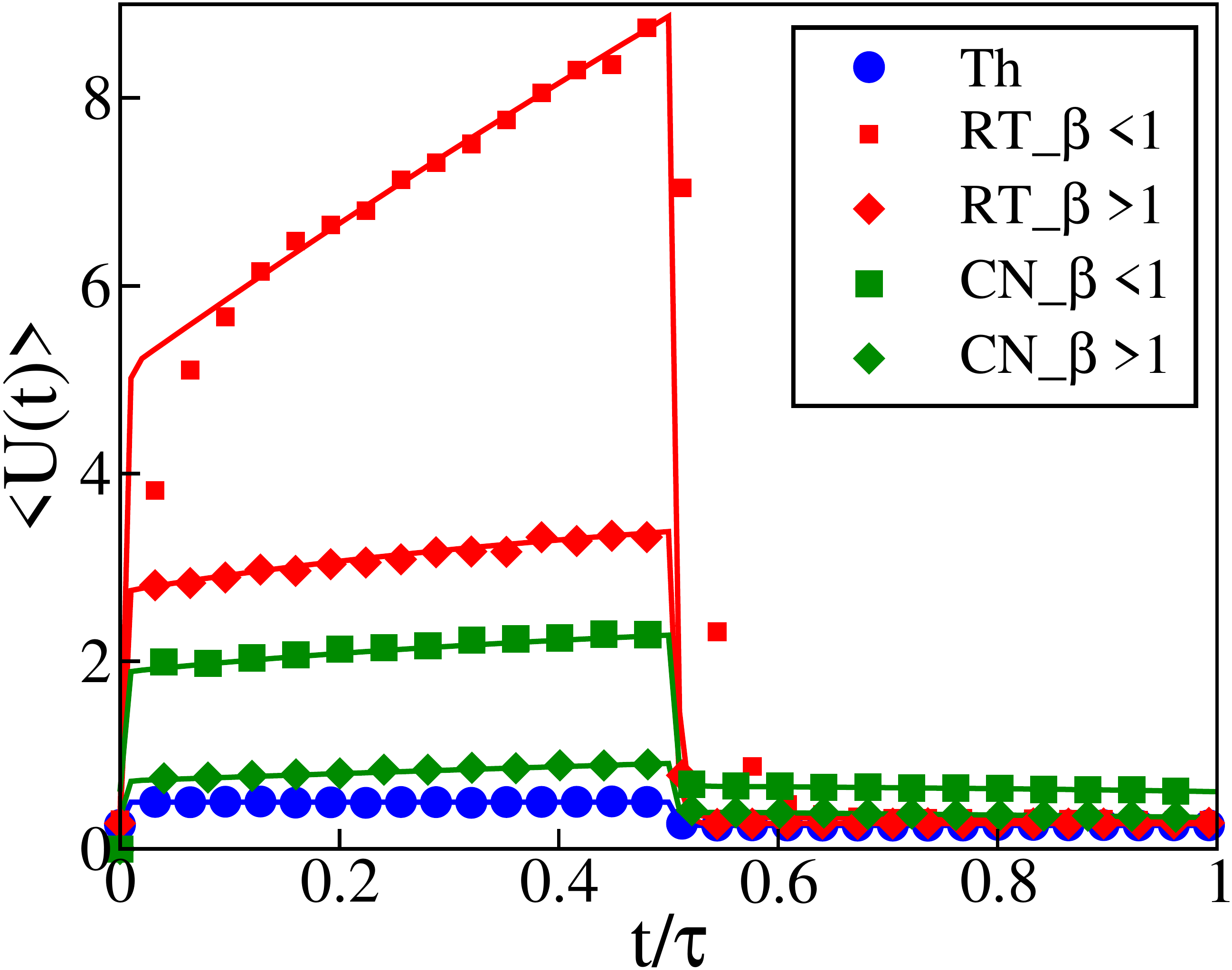}
	\caption{Internal energy as a function of the reduced cycle time $t/\tau$, for $\beta > 1$ and $\beta < 1$ for large $\tau$. We see that the RT model doesn't have a fixed $T^{eff}_{RT}$ when $\beta <1$. The parameters are $k_{1}=4.0$, $ k_{2}=2.0$, $\tau=100$, $\gamma_{1}=1.0$, $\gamma_{2}=2.0$, $T_{1}=1.0$, $T_{2}=0.5$, $A_{1}^+=9.0$, $A_{2}^+=1.0$, $D_{1}=5.0$, $D_{2}=2.0$, $\lambda_{1}=\lambda_{2}=0.4$ and $\tau_{a}=2.5$ (for $\beta<1$), $\lambda_{1}=\lambda_{2}=5.0$ and $\tau_{a}=0.2$ (for $\beta>1$).}
	\label{utvstdiffbeta} \end{figure}

We plot the average energy $\langle U(t)\rangle$ as a function of cycle time in Fig. \ref{utvstdiffbeta}. We observe that for the RT model, with $\beta>1$, the graph has similar nature as the passive model. However, for $\beta<1$,  $\langle U(t)\rangle$ increases rapidly with time, and after the half of the cycle, it decreases with time to a lower value. Therefore, depending on $\beta$, the behaviour of average internal energy within a cycle can be considerably different and a $T^{eff}_{RT}$ may not be defined in either halves of the cycle or $RT$  model requires even larger values of $\tau$ to show the $T^{eff}_{RT}$ like behavior. In contrast to RT dynamics, the CN model, due to its Gaussian nature, behaves quite similar to the passive model, except a minute increase in energy along the first half of the cycle and an overall higher $T^{eff}_{CN}$ in both halves of the cycle. This happens irrespective of whether $\beta >1$ or $\beta <1$. But the values of the $T^{eff}_{CN}$ can be considerably different depending on $\beta$. 
\begin{figure}[!t]
	\hspace{-1.5cm}
	\includegraphics[width=7.5cm,height=6.0cm,angle=0]{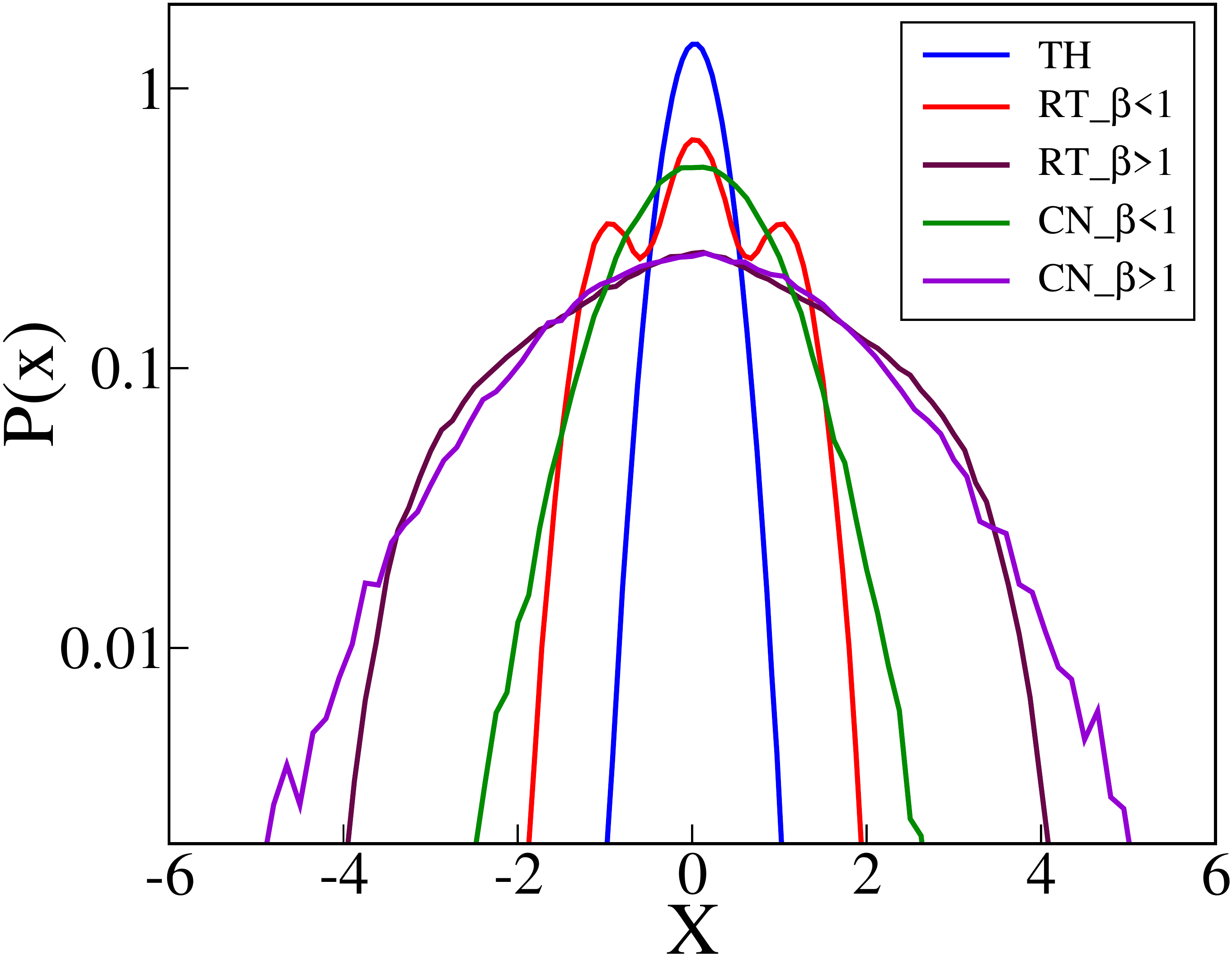}~~~
    \includegraphics[width=7.5cm,height=6.0cm,angle=0]{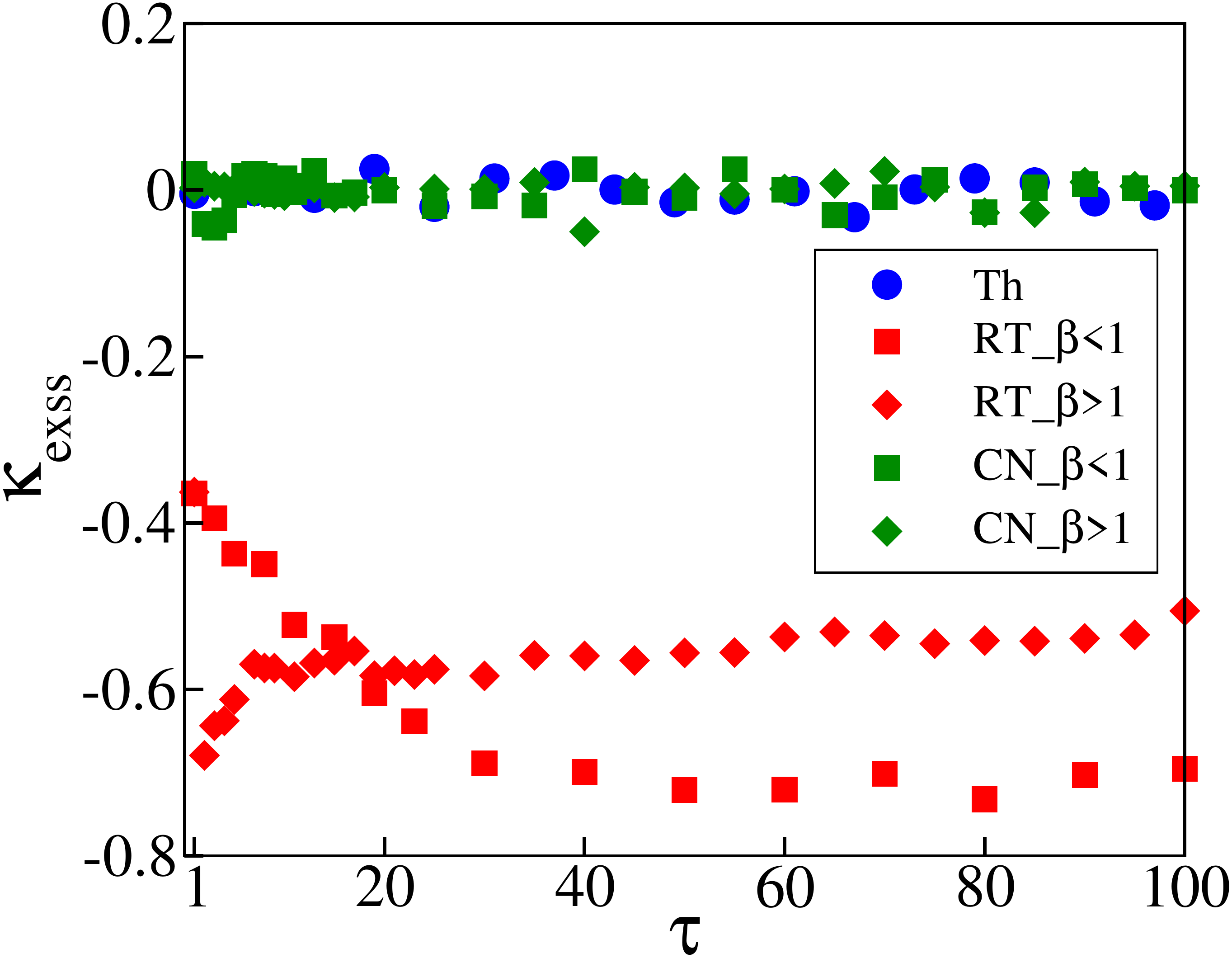}
	\caption{Log-lin plot of (Left panel) P(x) with x at time $\frac{\tau}{4}$, and $\kappa_{exss}$ with  $\tau$ (Right panel). We see that for RT model, the distributions are clearly non-Gaussian with larger widths than the passive case. For CN the distributions are Gaussian but widths are still larger than the passive case. The parameters are $k_{1}=4.0$, $ k_{2}=2.0$, $\tau=100$, $\gamma_{1}=1.0$, $\gamma_{2}=2.0$, $T_{1}=0.2$, $T_{2}=0.1$, $D_{1}=10.0$, $D_{2}=4.0$, $\lambda_{1}=\lambda_{2}=0.4$ and $\tau_{a}=2.5$ (for $\beta<1$), $\lambda_{1}=\lambda_{2}=5.0$ and $\tau_{a}=0.2$ (for $\beta>1$) and  $A_{1}^+=\sqrt{\frac{2D_1}{\tau_a}}$, $A_{2}^+=\sqrt{\frac{2D_2}{\tau_a}}$.}
	\label{pxdiffbeta}  
\end{figure}
\begin{figure}[!htbp]
	\hspace{-1cm}
	\includegraphics[width=7.5cm,height=6.0cm,angle=0]{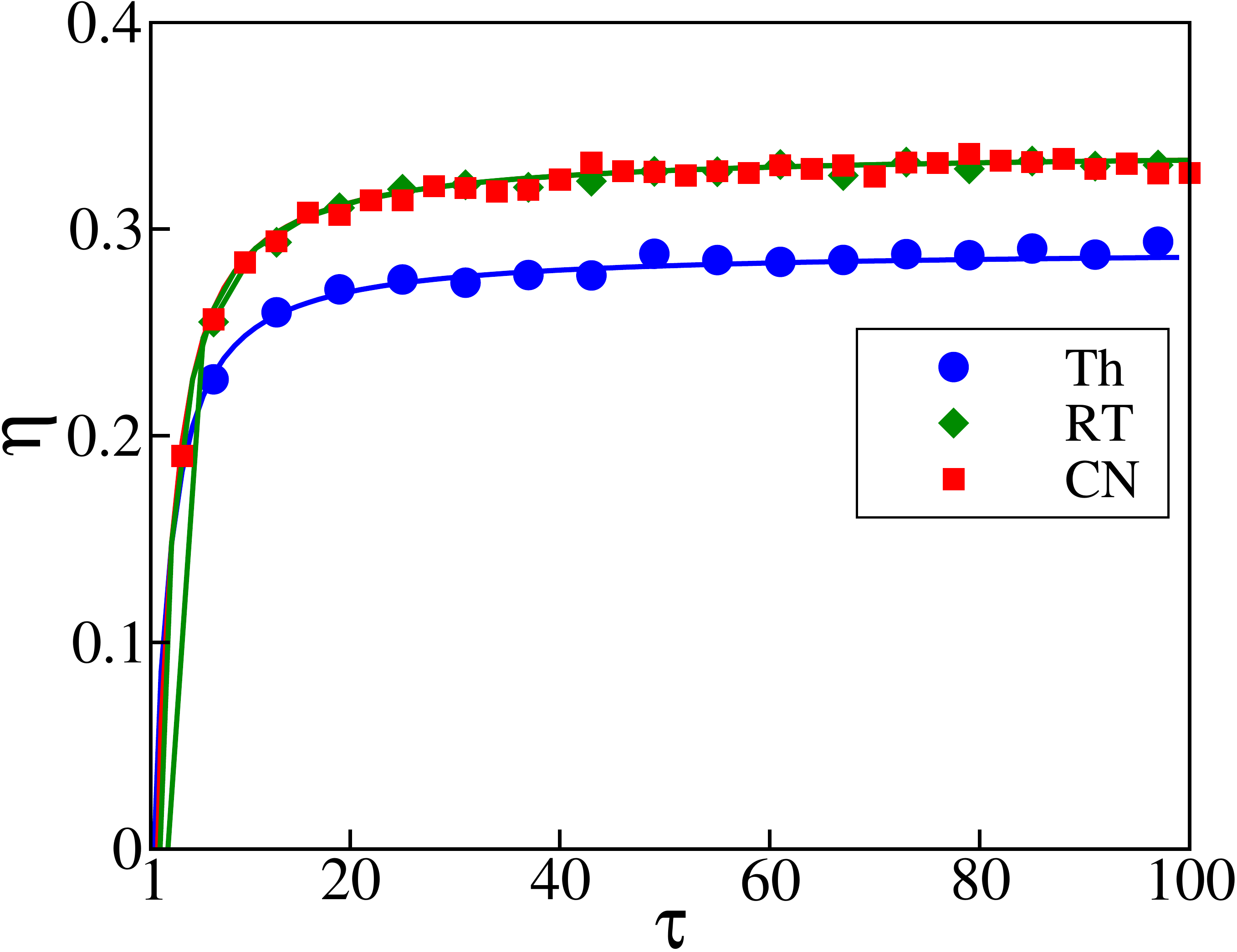}~~~
    \includegraphics[width=7.5cm,height=6.0cm,angle=0]{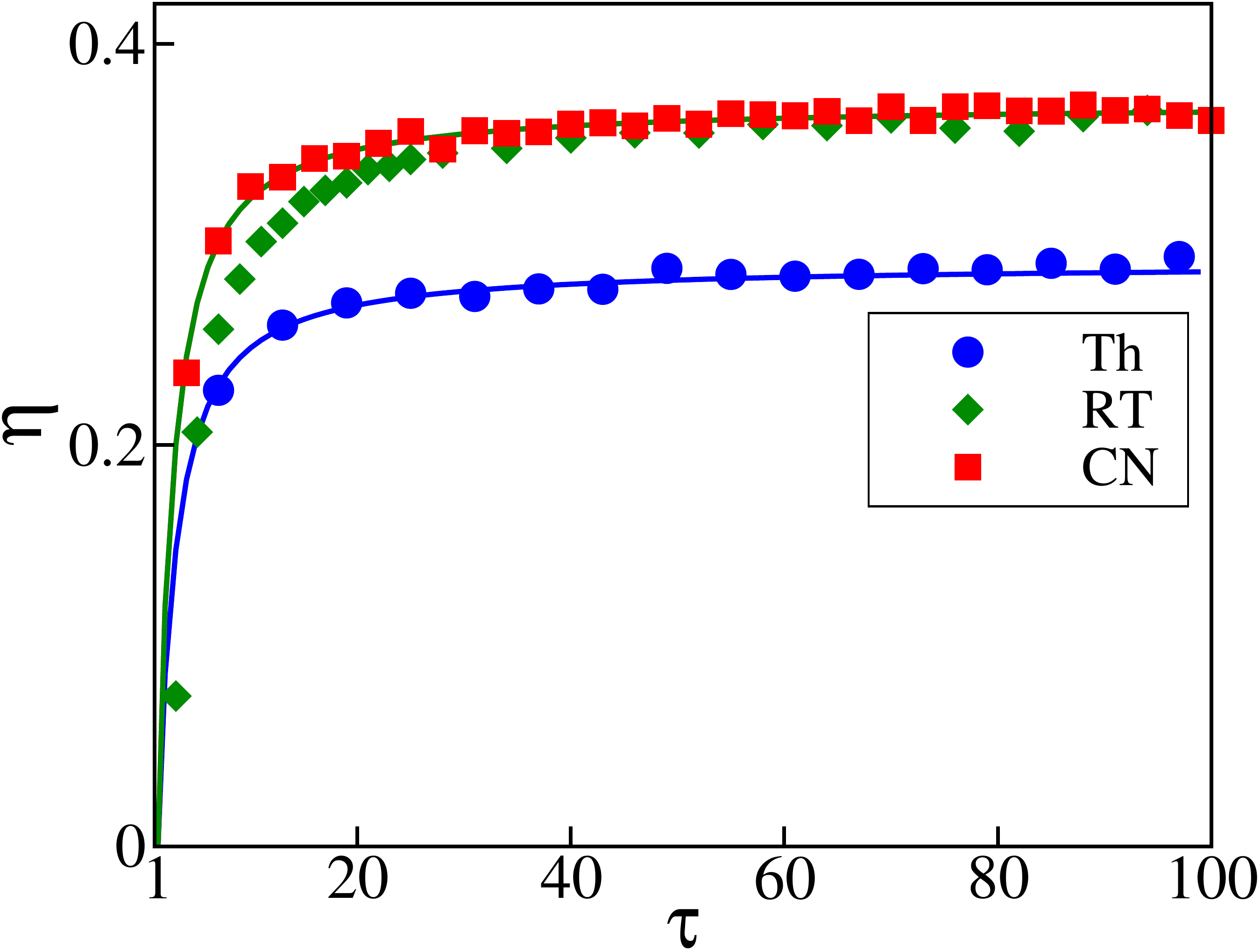}
	\caption{Efficiency $\eta$ as a function of the cycle time $\tau$. (Left panel) for $\beta <1$. (Right panel) for $\beta >1$. We see that for these parameters RT and CN engines always perform better than the passive engine. For new images The parameters are $k_{1}=4.0$, $ k_{2}=2.0$, $\tau=100$, $\gamma_{1}=1.0$, $\gamma_{2}=2.0$, $T_{1}=1.0$, $T_{2}=0.5$, $A_{1}^+=9.0$, $A_{2}^+=1.0$, $\lambda_{1}=\lambda_{2}=0.4$ and $\tau_{a}=2.5$ (for $\beta<1$),  $\lambda_{1}=\lambda_{2}=5.0$ and $\tau_{a}=0.2$ (for $\beta>1$) and  $D_{1}=\frac{(A_1^+)^{2}\tau_a}{2}$, $D_{2}=\frac{(A_2^+)^{2}\tau_a}{2}$.} 
	\label{etadiffbeta}  
\end{figure}

We also plot the steady state distribution of the positions in Fig. \ref{pxdiffbeta} for different values of $\beta$ at times $t=\tau/4$ (left panel) as discussed earlier. We observe that for passive engine and CN model, the distributions are clearly Gaussian as expected however for RT they are clearly non-Gaussian. The $\kappa_{exss}$ plot for different cases in Fig. \ref{pxdiffbeta} (right panel) is consistent to this observation. In active engine cases for $\beta <1$ or $>1$ the widths of the distributions for RT model are considerably broader than the thermal noise giving rise to larger efficiencies than the passive engine as seen in Fig. \ref{etadiffbeta} (left and right panel). On the other hand the efficiencies of both RT and CN model match if the parameters are chosen appropriately (Fig. \ref{etadiffbeta} (left panel)) as expected. This reflects that the performance of the active micro heat engines considered here, is enhanced in comparison to their passive counterpart due to their non-Markovian feature attributed to their activity.      

\section{Conclusion}
In this paper we have studied different models of active heat engines and compared them with the thermal or passive engine. Unlike earlier studies, in case of passive engine we can calculate all the thermodynamic quantities of interests completely analytically. This is done by choosing a particular form of the engine protocol. The two active engines that we have studied differ in the nature of the active noises. The noise in the RT model is highly non-Gaussian and persistent but in the CN model it is Gaussian and persistent. However, both the models drive the system out-of-equilibrium (even for a constant trap strength) as the active noises do not follow fluctuation-dissipation relation. We observe that the behavior of the engines depend quite crucially on the parameters, especially the non-Gaussian distributions of position characterized by $\kappa_{exss}$ and more importantly $\beta$ that measures the competition between the persistence in the active noise and the strength of harmonic confinement. Thus, the efficiency of the active engines can be made better than that of the passive engine. We also note that the behavior of the active engines (especially, the position distributions) depend quite sensitively on the strength of the thermal noise which is present in both the active engines. If the strength of thermal noise is even slightly higher, then the non-Gaussian nature of the distributions (trimodal structure in Fig. \ref{pxdiffbeta}) is lost. Hence, looking at just the distributions and the second moments of the particle position may not suffice in such cases. We believe our detailed study of different active engines try to answer important questions, like under which conditions enhancement in the efficiency from passive engines can be achieved? which are the important parameters that characterize a given active-engine? \cite{Lee2021}. Moreover, the external protocol used in the exactly solved model for the passive engine discussed here can be easily implemented in an experiment and compared with their active counterparts.

\section{acknowledgment}
RaM gratefully acknowledges Science and Engineering Research Board (SERB), India for financial support through the MATRICS grant (No. MTR/2020/000349). AS thanks the start-up grant from University Grants Commission (UGC) via UGC Faculty recharge program (UGCFRP) and the Core Research Grant (CRG/2019/001492) from SERB, India. RiM and RaM acknowledge Sandwich Training Educational Programme (STEP) by Abdus Salam International Center for Theoretical Physics (ICTP), Trieste, Italy where a part of this work was done. We also thank \'{E}dgar Rold\'{a}n and Gonzalo Manzano Paule for useful discussions. We dedicate this work to our friend, teacher and collaborator Prof. Arun Jayannavar, Institute of Physics, Bhubaneshwar, India, who passed away recently. Arun you will be missed dearly.

\end{document}